\documentclass[twocolumn]{aastex63}

\def\cii{\hbox{{\rm [C {\scriptsize II}]}}}

\def\13cii{\hbox{{\rm [$^{13}$C {\scriptsize II}]}}}

\def\oi{\hbox{{\rm [O {\scriptsize I}]}}}
\def\hii{\hbox{{\rm H {\scriptsize II}}}}

\newcommand{\um}{$\mu$m}
\newcommand{\herschel}{\textit{Herschel}}
\newcommand{\planck}{\textit{Planck}}

\defcitealias{MSP1991}{MSP}
\defcitealias{TFT2007}{TFT07}
\defcitealias{VA2013}{VA13}
\defcitealias{VPHAS_Wd2_2015}{VPHAS+}

\usepackage{gensymb}
\usepackage[flushleft]{threeparttable}
\usepackage{amsmath}

\received{December 8, 2020}
\accepted{April 8, 2021}
\submitjournal{ApJ}

\shorttitle{RCW~49}
\shortauthors{Tiwari et al.}

\graphicspath{{./}{figures/}}

\begin{document}

\title{SOFIA FEEDBACK survey: exploring the dynamics of the stellar wind driven shell of RCW~49}

\author{M. Tiwari}
\affiliation{University of Maryland, Department of Astronomy, College Park, MD 20742-2421, USA}
\affiliation{Max-Planck Institute for Radioastronomy, Auf dem H\"{u}gel, 53121 Bonn, Germany}

\author{R. Karim}
\affiliation{University of Maryland, Department of Astronomy, College Park, MD 20742-2421, USA}

\author{M. W. Pound}
\affiliation{University of Maryland, Department of Astronomy, College Park, MD 20742-2421, USA}

\author{M. Wolfire}
\affiliation{University of Maryland, Department of Astronomy, College Park, MD 20742-2421, USA}

\author{A. Jacob}
\affiliation{Max-Planck Institute for Radioastronomy, Auf dem H\"{u}gel, 53121 Bonn, Germany}

\author{C. Buchbender}
\affiliation{I. Physik. Institut, University of Cologne, Z\"{u}lpicher Str. 77, 50937 Cologne, Germany}

\author{ R. G\"{u}sten}
\affiliation{Max-Planck Institute for Radioastronomy, Auf dem H\"{u}gel, 53121 Bonn, Germany}

\author{C. Guevara}
\affiliation{I. Physik. Institut, University of Cologne, Z\"{u}lpicher Str. 77, 50937 Cologne, Germany}

\author{R.D. Higgins}
\affiliation{I. Physik. Institut, University of Cologne, Z\"{u}lpicher Str. 77, 50937 Cologne, Germany}

\author{S. Kabanovic}
\affiliation{I. Physik. Institut, University of Cologne, Z\"{u}lpicher Str. 77, 50937 Cologne, Germany}

\author{C. Pabst}
\affiliation{Leiden Observatory, Leiden University, PO Box 9513, 2300 RA Leiden, Netherlands.}

\author{O. Ricken}
\affiliation{Max-Planck Institute for Radioastronomy, Auf dem H\"{u}gel, 53121 Bonn, Germany}

\author{N. Schneider}
\affiliation{I. Physik. Institut, University of Cologne, Z\"{u}lpicher Str. 77, 50937 Cologne, Germany}

\author{R. Simon}
\affiliation{I. Physik. Institut, University of Cologne, Z\"{u}lpicher Str. 77, 50937 Cologne, Germany}

\author{J. Stutzki}
\affiliation{I. Physik. Institut, University of Cologne, Z\"{u}lpicher Str. 77, 50937 Cologne, Germany}

\author{A. G. G. M. Tielens}
\affiliation{University of Maryland, Department of Astronomy, College Park, MD 20742-2421, USA}
\affiliation{Leiden Observatory, Leiden University, PO Box 9513, 2300 RA Leiden, Netherlands.}

\begin{abstract}

We unveil the stellar wind driven shell of the luminous massive star-forming region of RCW~49 using SOFIA FEEDBACK observations of the \cii\,~158~$\mu$m line. The complementary dataset of the $^{12}$CO and $^{13}$CO $J$ = 3 $\to$ 2 transitions is observed by the APEX telescope and probes the dense gas toward RCW~49. Using the spatial and spectral resolution provided by the SOFIA and APEX telescopes, we disentangle the shell from a complex set of individual components of gas centered around RCW~49. We find that the shell of radius $\sim$ 6~pc is expanding at a velocity of 13~km~s$^{-1}$ toward the observer. Comparing our observed data with the ancillary data at X-Ray, infrared, sub-millimeter and radio wavelengths, we investigate the morphology of the region. The shell has a well defined eastern arc, while the western side is blown open and is venting plasma further into the west. Though the stellar cluster, which is $\sim$ 2~Myr old gave rise to the shell, it only gained momentum relatively recently as we calculate  the shell's expansion lifetime $\sim$ 0.27~Myr, making the Wolf-Rayet star WR20a a likely candidate responsible for the shell's re-acceleration.
 
\end{abstract}



\section{Introduction} \label{sec:intro}

\begin{figure*}
\centering
\includegraphics[width=175mm]{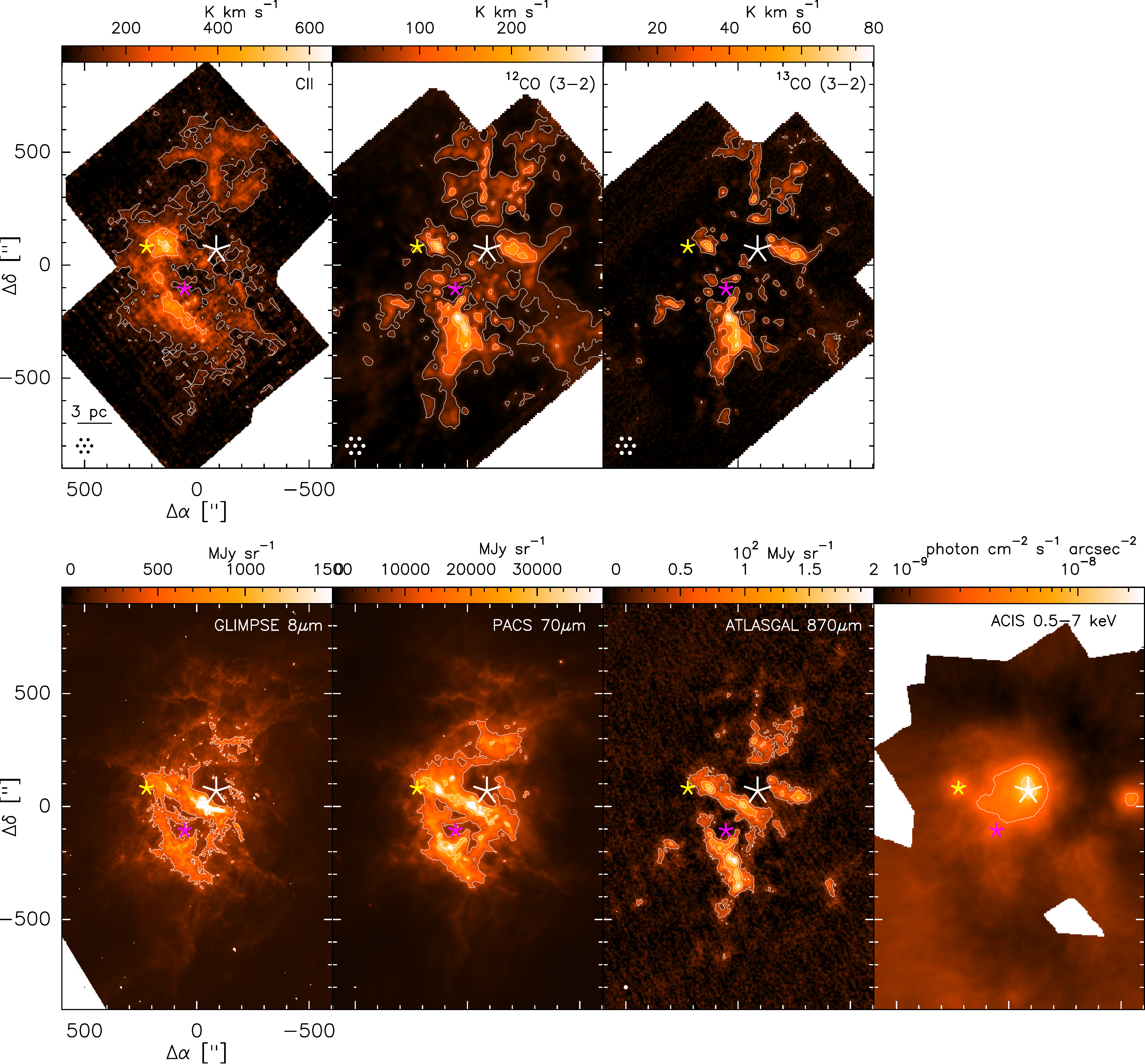}

\caption{Top panel: velocity integrated intensity maps in the range of -25 to 30~km~s$^{-1}$, from left to right: of \cii\ $^2$P$_{3/2}$ $\to$ $^2$P$_{1/2}$ fine-structure line, $J$ = 3 $\to$ 2 transitions of $^{12}$CO and $^{13}$CO. The maps are in original resolution with their original beam sizes and the receiver array geometry is shown in the bottom left of each panel. Bottom panel: Emission images, from left to right, of the 8~$\mu$m GLIMPSE, 70~$\mu$m PACS, 870~$\mu$m ATLASGAL and ACIS data toward RCW~49. The center of Wd2 cluster at R.A.($\alpha$, J2000) = 10$^{h}$24$^{m}$11$^{s}$.57 and Dec.\,($\delta$, J2000) = $-$57$\degree$46$\arcmin$42.5$\arcsec$ is marked with a white asterisk. The OV5 star and the WR20b star are marked with yellow and pink asterisks, respectively. For \cii\ emission, the contour levels are smoothed to a pixel size of 15$\arcsec$ and are 20\% to 100\% in steps of 20\% of the corresponding peak emission. For the presented maps, the peak emission for \cii\,, $^{12}$CO and $^{13}$CO are 650~K~km~s$^{-1}$, 300~K~km~s$^{-1}$ and 80~K~km~s$^{-1}$, respectively. For $^{12}$CO emission, the contour levels are 10\% to 100\% in steps of 20\% of the corresponding peak emission. For $^{13}$CO and 870 $\mu$m emission, the contour levels are 15\% to 100\% in steps of 20\% of the corresponding peak emission. For 8~$\mu$m emission, the contour levels are 15\% to 100\% in steps of 30\% of the corresponding peak emission. For 70~$\mu$m, the contour levels are 20\% to 100\% in steps of 30\% of the corresponding peak emission. For 0.5--7~keV emission, the contour levels are 10\% to 100\% in steps of 30\% of the corresponding peak emission. \label{vel-int-maps}}
\end{figure*}

One of the most important problems in modern astrophysics is to understand the role of massive stars in driving various physical and chemical processes in the interstellar medium (ISM). Massive stars inject an immense amount of mechanical and radiative energy into their immediate vicinity. Stellar winds are responsible for the mechanical energy input, which can push the gas into shell-like structures (as in the Rosette Nebula, \citealt{2018MNRAS.475.3598W} and in the Orion Nebula, \citealt{2019Natur.565..618P}). The radiative energy input comes from the heating of gas through stellar extreme-ultraviolet (EUV, $h\nu$ $>$ 13.6~eV) and far-UV (FUV, 6 $<$ $h\nu$ $<$ 13.6~eV) photons that can ionize atoms, dissociate molecules and heat the gas giving rise to \hii\ regions and photodissociation regions (PDRs). These stellar feedback mechanisms power the expansion of \hii\ regions and shock fronts causing morphological features that appear as shells or bubbles in the ISM. Observational studies at near-infrared (IR) wavelengths led \citet{2006ApJ...649..759C} to report that these features are ubiquitous in our Galaxy and the authors coined the term ``bubbles'' by stating ``We postulate that the rings are projections of three-dimensional shells and henceforth refer to them as bubbles''. Processes that cause these shells can disrupt molecular clouds, thereby halting star formation or can compress the gas at the edges of the \hii\ regions, triggering star formation (\citealt{1977ApJ...214..725E}, \citealt{1997ApJ...476..166W} and \citealt{2010A&A...518L.101Z}). Thus, shells are ideal laboratories to study positive and negative feedback generated by massive star formation. Furthermore, in order to trace these shells, we can use the 1.9~THz fine-structure line of ionised carbon, C$^+$ (\cii\,), which is one of the major coolants of the ISM and also among the brightest lines in PDRs (\citealt{1972ARA&A..10..375D}, \citealt{1991ApJ...373..423S}, \citealt{1994ApJ...434..587B}). As the ionization potential (11.3~eV) of carbon (C) is less than that of hydrogen (H) (13.6~eV), \cii\ traces the transition from H$^+$ to H and H$_2$ \citep{1999RvMP...71..173H}. While the \cii\ layer probes warm atomic gas from the surface of clouds at low visual extinction ($A_{\rm v}$ $\lessapprox$ 4 magnitude), the rotational transitions of CO probe cooler molecular gas at larger $A_{\rm v}$ \citep{1999RvMP...71..173H} deeper into the cloud clumps. \\

RCW~49 is among the most luminous and massive star forming regions of the southern Galaxy located close to the tangent of the Carina arm at $l$ = 284.3$\degree$, $b$ = -0.3$\degree$.
The earlier heliocentric distance measurements to RCW~49 varied from 2 to 8~kpc, as discussed in \citet{2004ApJS..154..322C}, \citet{Rauw2007}, and \citet{2009ApJ...696L.115F}.
\citet{Drew2018Wd2}, in their discussion of the distance, note that photometric studies of the stars powering RCW~49 have more recently tended towards 4--6~kpc rather than the 2--8~kpc span that is usually quoted, although \citet{Rauw2007, 2011A&A...535A..40R} maintain that the distance must be 8~kpc.
Three of the most recent works determine a distance of about 4.2~kpc \citep{VA2013, 2015AJ....150...78Z, 2018A&A...618A..93C}.
\citet{VA2013} and \citet{2015AJ....150...78Z} in particular conducted photometric studies with two independent sets of observations and agreed on a $\sim$4~kpc distance.
This is consistent with the 4.2~kpc \textit{Gaia} parallax distance reported by \citet{2018A&A...618A..93C}, though \citet{Drew2018Wd2} point out that the uncertainties on these small parallax values are considerable.
In accordance with the photometric study of \citet{VA2013} and the consistent measurement by \citet{2015AJ....150...78Z}, we adopt a distance of 4.16~kpc.

RCW~49 
contains a bright \hii\ region ionized by a compact stellar cluster, Westerlund~2 (Wd2), comprising 37 OB stars and $\sim$ 30 early type OB star candidates around it
\citep{TFT2007, Ascenso2007, 2011A&A...535A..40R, VPHAS_Wd2_2015, 2015AJ....150...78Z}.
There is a binary Wolf-Rayet star (WR20a) associated with the central Wd2 cluster, suggested to be one of the most massive binaries in the Galaxy \citep{Rauw2005}.
RCW~49 also hosts an O5V star and another Wolf-Rayet star (WR20b), both a few arcminutes away from the geometrical cluster center.
These and a handful of other massive stars in the cluster periphery may have been ejected from Wd2 \citep{Drew2018Wd2}.
Age estimates generally suggest that the cluster is not much older than 2~Myr \citep{Ascenso2007, 2015AJ....150...78Z}. \\

\begin{figure*}
\centering
\includegraphics[width=160mm]{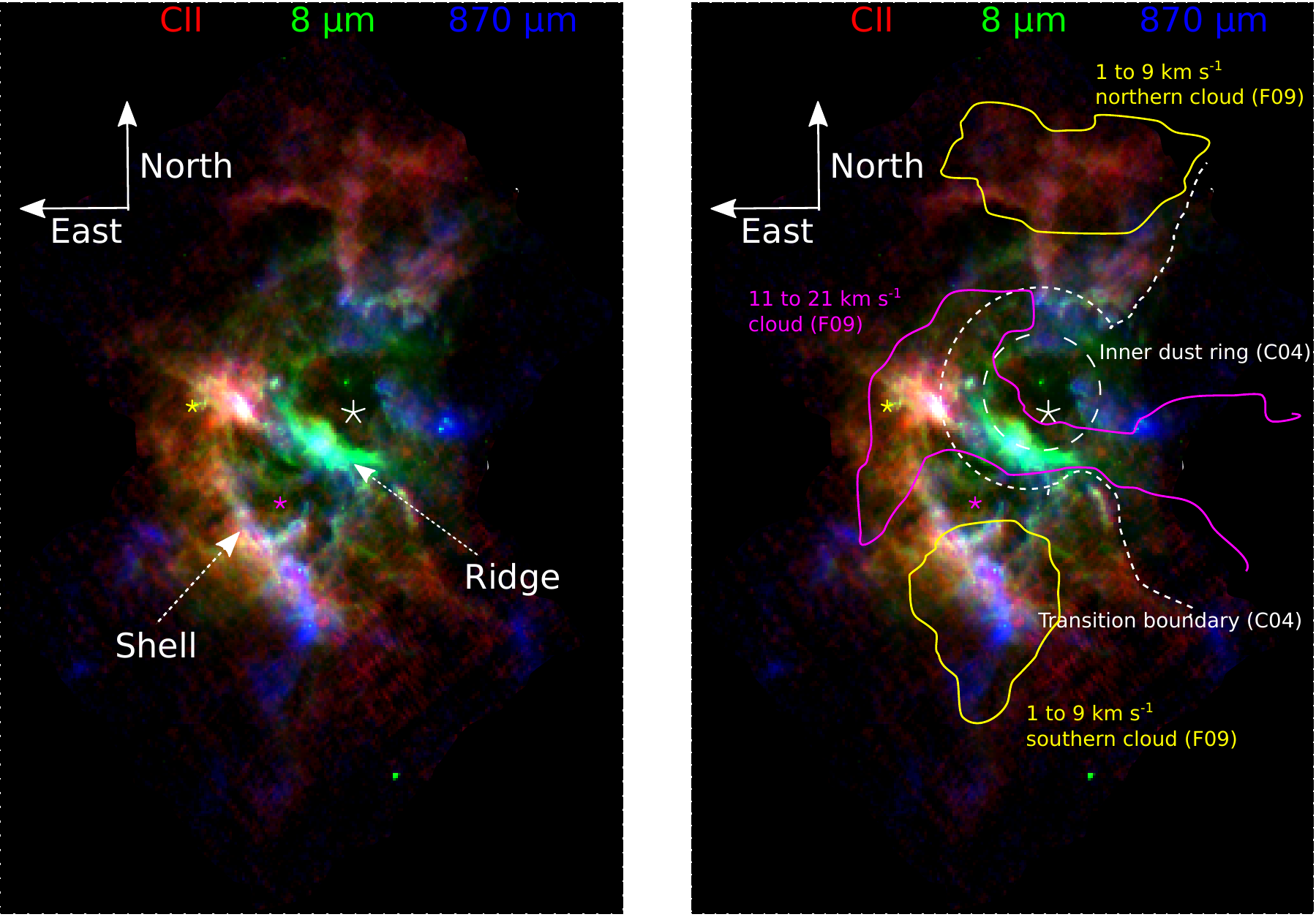}

\caption{RGB images of \cii\ (red), GLIMPSE 8~$\mu$m (green) and ATLASGAL 870~$\mu$m (blue) emission toward RCW~49. The Wd2 cluster's center, the OV5 and the WR20b stars are marked with white, yellow and pink asterisks, respectively. The ridge and the shell are marked in the left panel, while the inner dust ring (white dashed circle) and the transition boundary (white dotted line) are marked similar to \citet[Fig.~1]{2004ApJS..154..322C} in the right panel. The $^{12}$CO clouds as shown in \citet[Fig.~1~(a) and (c)]{2009ApJ...696L.115F} are also outlined. In yellow are the northern and southern blobs of the cloud within the velocity range of 1 to 9~km~s$^{-1}$ and in magenta is the cloud within the velocity range of 11 to 21~km~s$^{-1}$. These clouds are discussed more in Sections 3.2, 3.3 and 3.4.}

\label{rgb-cii-8-870} 
\end{figure*}

In this paper, we report one of the first results of the  Stratospheric Observatory For Infrared Astronomy (SOFIA, \citealt{2012ApJ...749L..17Y}) legacy program FEEDBACK\footnote{https://feedback.astro.umd.edu} \citep{2020PASP..132j4301S} performed with the SOFIA and the Atacama Pathfinder Experiment (APEX\footnote{APEX, the Atacama Pathfinder Experiment is a collaboration between the Max-Planck-Institut f\"{u}r Radioastronomie, Onsala Space Observatory (OSO), and the European Southern Observatory (ESO).}, \citealt{2006A&A...454L..13G}). The FEEDBACK Legacy Program was initiated to quantify the mechanical and radiative feedback of massive stars on their environment.  A wide range of sources were selected to be observed that allow a systematic survey of the effects of different feedback mechanisms due to star formation activity (with a single O type star, groups of O type stars, compact clusters, mini starbursts, etc.), morphology of their environment, and the evolutionary stage of star formation. In particular, RCW~49 was selected to study the feedback of the compact stellar cluster, Wd2 and the Wolf-Rayet stars on their surrounding molecular clouds.

We use the fine-structure line of \cii\ to probe the shell associated with RCW~49 and disentangle its dense gas component using the CO observations. We quantify the stellar wind feedback responsible for the evolution of the shell of RCW~49 and describe the shell's morphology. In Sect.~2, we describe the observations. The qualitative and quantitative analysis of the data are reported in Sect.~3. Both small and large scale effects of the stellar feedback in RCW~49 are discussed in Sect.~4 and the results of this study are summarised in Sect.~5.

\section{Observations}

\subsection{SOFIA Observations}

The \cii\ line at 1.9~THz was observed during three flights from Christchurch, New Zealand on 7th, 10th, and 11th of June 2019, using upGREAT\footnote{German Receiver for Astronomy at Terahertz. (up)GREAT is a development by the MPI f\"ur Radioastronomie and the KOSMA/Universit\"at zu K\"oln, in cooperation with the DLR Institut f\"ur Optische Sensorsysteme.} \citep{2018JAI.....740014R}. upGREAT consists of a 2 $\times$ 7 pixel low-frequency array (LFA) that was tuned to the \cii\ line, and in parallel a seven pixel high frequency array (HFA) that was tuned to the \oi\ 63 $\mu$m line. Both arrays observe in parallel, but here, we
only present the \cii\ data.  The half-power beam widths are 14.1$\arcsec$ (1.9~THz) and 6.3$\arcsec$ (4.7~THz), determined by the instrument and telescope optics, and confirmed by observations of planets. The final pixel size of the \cii\ map is 7.5$\arcsec$. The observation region was split into 12 individual `tiles', each covering an area of (7.26~arcmin)$^{2}$ = 52.7~arcmin$^2$. During the three flights, eight tiles were observed ($\sim$66\% of the planned area). Each tile was covered four times and they were tilted 40$\degree$ against the R.A.
axis (counter clockwise against North) for horizontal scans and perpendicular to that for the corresponding vertical scans. As a consequence of its hexagonal geometry, the array was rotated by 19$\degree$ against the tile scan direction to achieve equal spacing between the on-the-fly scan lines. The gaps between the pixels are approximately two beam widths (31.7$\arcsec$ for the LFA and 13.8$\arcsec$ for the HFA), which result into a projected pixel spacing of 10.4$\arcsec$ for the LFA and 4.6$\arcsec$ for the HFA after rotation (for more details, see the SOFIA Science Center's Planning observations webpage\footnote{https://www.sofia.usra.edu/science/proposing-and-observing/observers-handbook-cycle-9/6-great/62-planning-observations}). The second two coverages are then shifted by 36$''$ to achieve the best possible coverage for the \oi\ line.  All observations were carried out in the array-on-the-fly mapping mode. The map center was at 10$^h$24$^m$11$^s$.57, -57$\degree$46$\arcmin$42$\arcsec$.5 (J2000), the reference position at  10$^h$27$^m$17$^s$.42 -57$\degree$13$\arcmin$42$\arcsec$.60. For more observational and technical details, see \citet{2020PASP..132j4301S}. \\ 

As backend, a Fast Fourier Transform Spectrometer (FFTS)
with 4~GHz instantaneous bandwidth and a frequency resolution of 0.244~MHz was used \citep{2012A&A...542L...3K}. The \cii\ data thus have a native velocity
resolution of 0.04~km~s$^{-1}$. We use here data that was re-binned to a resolution of
0.2~km~s$^{-1}$. Spectra are presented on a main beam brightness
temperature scale $T_{\rm mb}$ with an average main beam efficiency of 0.65. The forward efficiency is $\eta_{\rm f}$ = 0.97.  From the spectra, a first order baseline was removed and the data-quality was improved by identifying and correcting systematic baseline features with a novel method for data reduction that makes use of a Principal Component Analysis (PCA) of reference spectra as described in Appendix~\ref{sec:pca_appen}.

\begin{figure*}[htp]
\includegraphics[width=180mm]{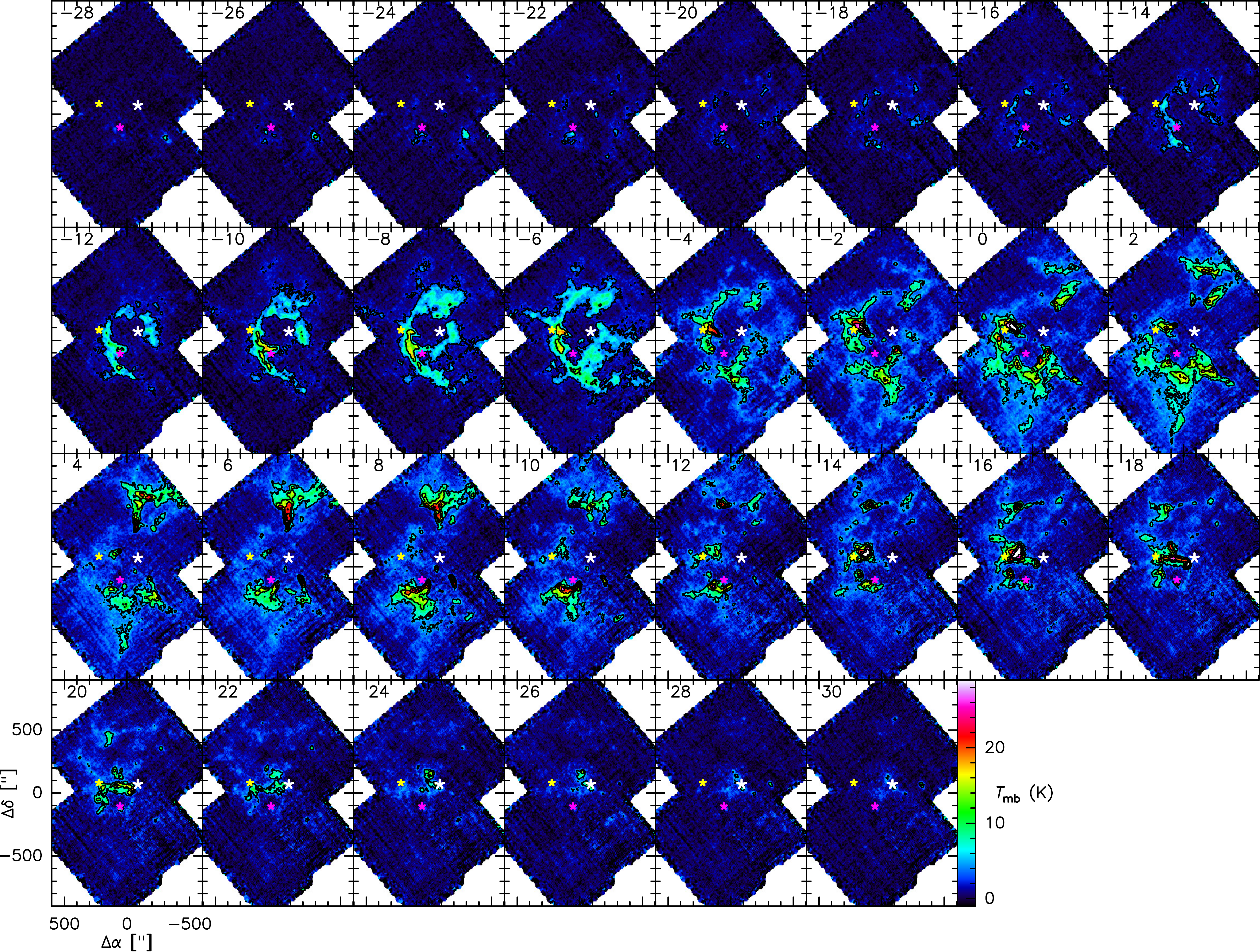}

\caption{Velocity channel maps of \cii\ emission toward RCW~49 with a channel width of 2~km~s$^{-1}$. The velocity (in km~s$^{-1}$) of each channel is shown in top left of each panel. The Wd2 cluster's center, the OV5 and the WR20b stars are marked with white, yellow and pink asterisks respectively.}  

\label{cii-chan} 
\end{figure*}

\begin{figure*}[htp]
\includegraphics[width=180mm]{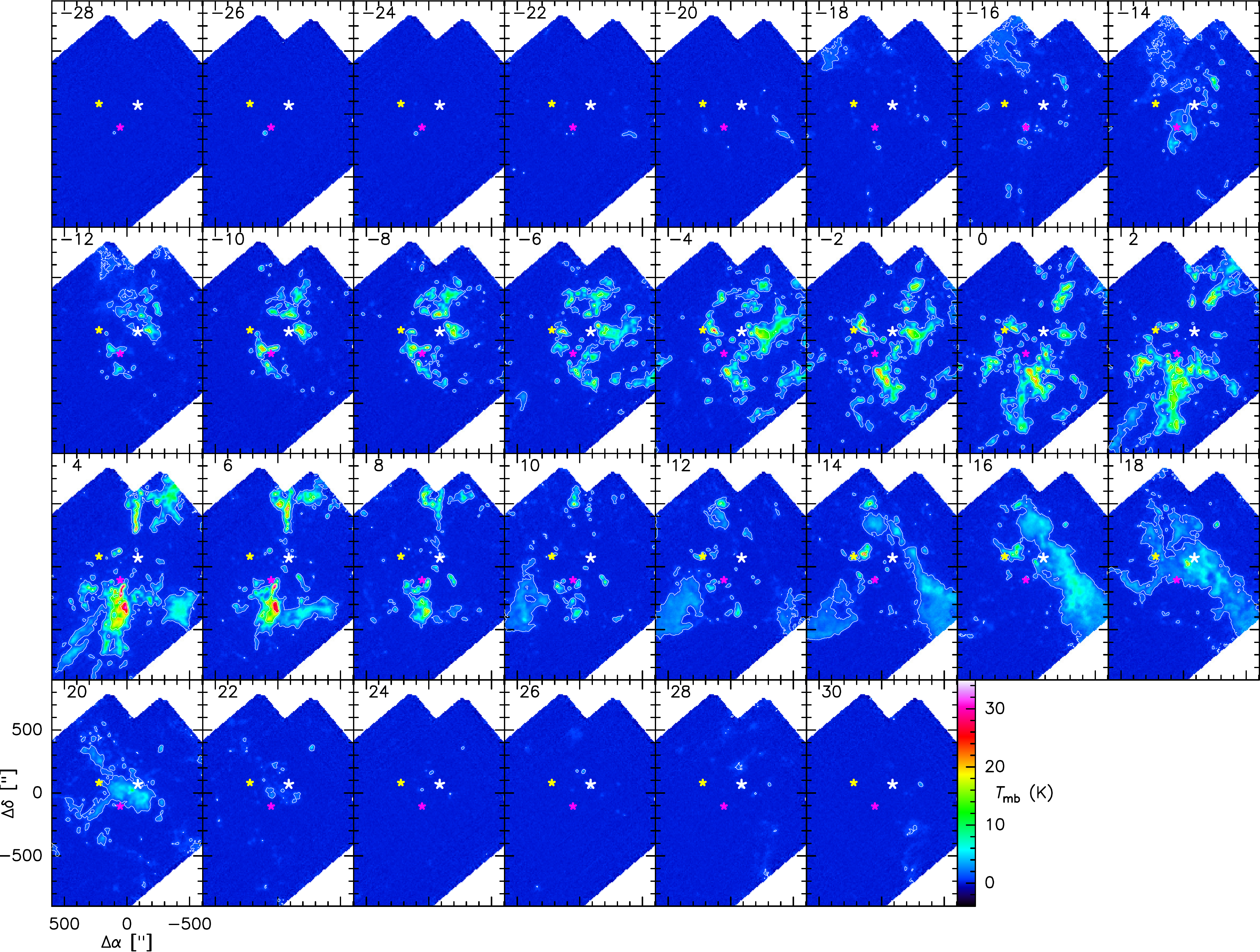}

\caption{Velocity channel maps of $^{12}$CO (3-2) emission toward RCW~49 with a channel width of 2~km~s$^{-1}$. The velocity (in km~s$^{-1}$) of each channel is shown in top left of each panel. The Wd2 cluster's center, the OV5 and the WR20b stars are marked with white, yellow and pink asterisks respectively.}  

\label{co-chan} 
\end{figure*}

\begin{figure*}[htp]

\includegraphics[width=180mm]{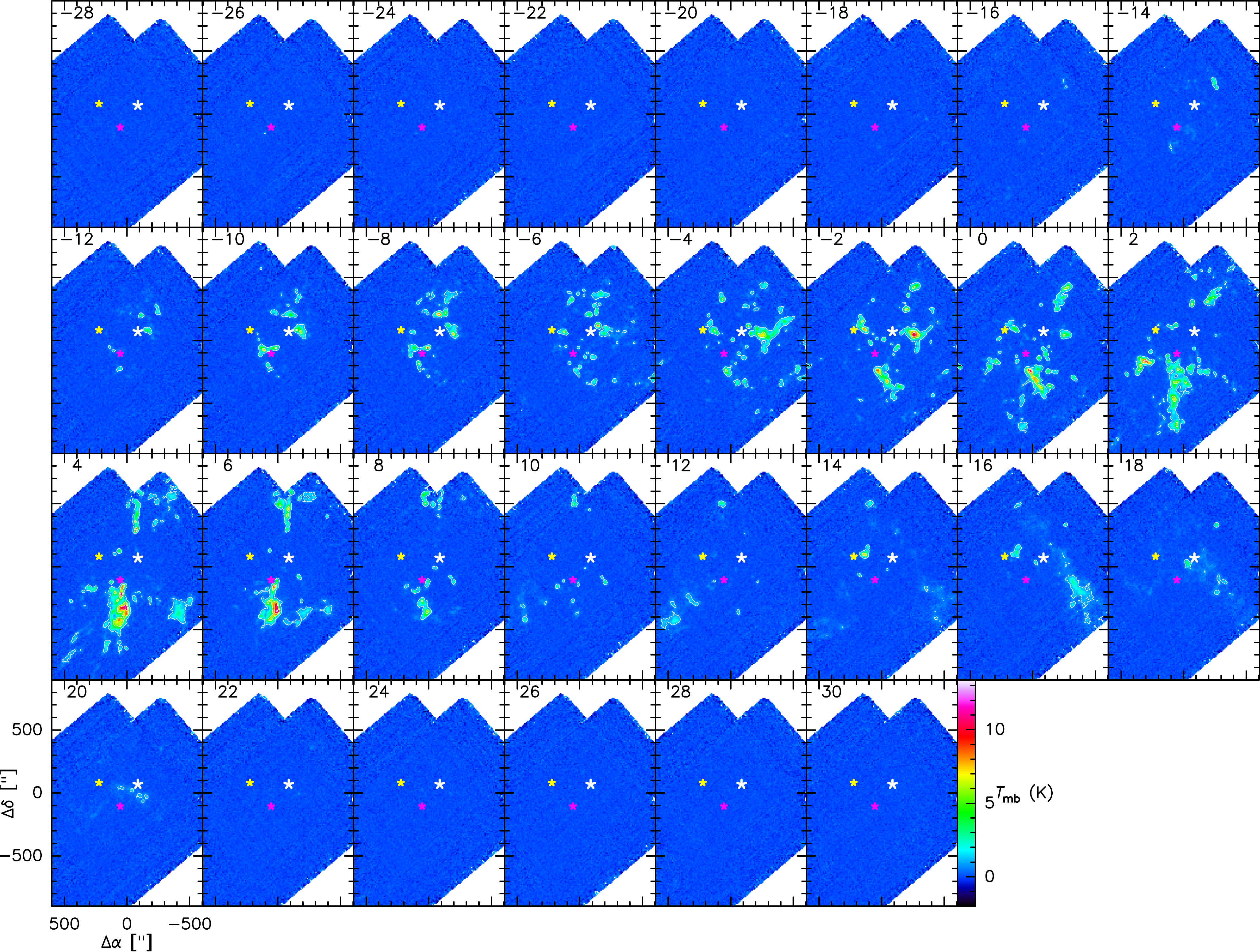}

\caption{Velocity channel maps of $^{13}$CO (3-2) emission toward RCW~49 with a channel width of 2~km~s$^{-1}$. The velocity (in km~s$^{-1}$) of each channel is shown in top left of each panel. The Wd2 cluster's center, the OV5 and the WR20b stars are marked with white, yellow and pink asterisks respectively.}  

\label{13co-chan} 
\end{figure*}

\subsection{APEX Observations}
RCW~49 was mapped on September 25-26, 2019 in good weather conditions (precipitable water vapor, pwv = 0.5 to 1~mm) in the $^{13}$CO(3-2) and $^{12}$CO(3–2) transitions using the LAsMA array on the APEX telescope \citep{2006A&A...454L..13G}. LAsMA is a 7-pixel single polarization heterodyne array that allows simultaneous observations of the two isotopomers in the upper ($^{12}$CO) and lower ($^{13}$CO) sideband of the receiver, respectively. The array is arranged in a hexagonal configuration around a central pixel with a spacing of about two beam widths ($\theta_{\rm mb}$ = 18.2$\arcsec$ at 345.8~GHz) between the pixels. It uses a K mirror as de-rotator. The backends are advanced FFTS \citep{2012A&A...542L...3K} with a bandwidth of 2 $\times$ 4~GHz and a native spectral resolution of 61~kHz.

The mapping was done in total power on-the-fly mode, with the scanning directions against N at -40$\degree$ and -130$\degree$ for the orthogonal scans, respectively. The map center and the reference position were the same as the SOFIA observations. The latter was verified to be free of CO emission at a level of $<$ 0.1~K. A total area of 570 arcmin$^2$ was observed, split into 4 tiles. Each tile was scanned with 6$\arcsec$ spacing (oversampled) in scanning direction, with a spacing of 9$\arcsec$ between rows, resulting in uniformly sampled maps with high fidelity. All spectra are calibrated in $T_{\rm mb}$ (main-beam efficiency $\eta_{\rm mb}$ = 0.68 at 345.8~GHz). A linear baseline was removed, all data resampled into 0.2~km~s$^{-1}$ spectral bins. The final data cubes are constructed with a pixel size of 9.5$\arcsec$ (the beam after gridding is 20$\arcsec$).

\begin{figure}[htp]

\includegraphics[width=85mm]{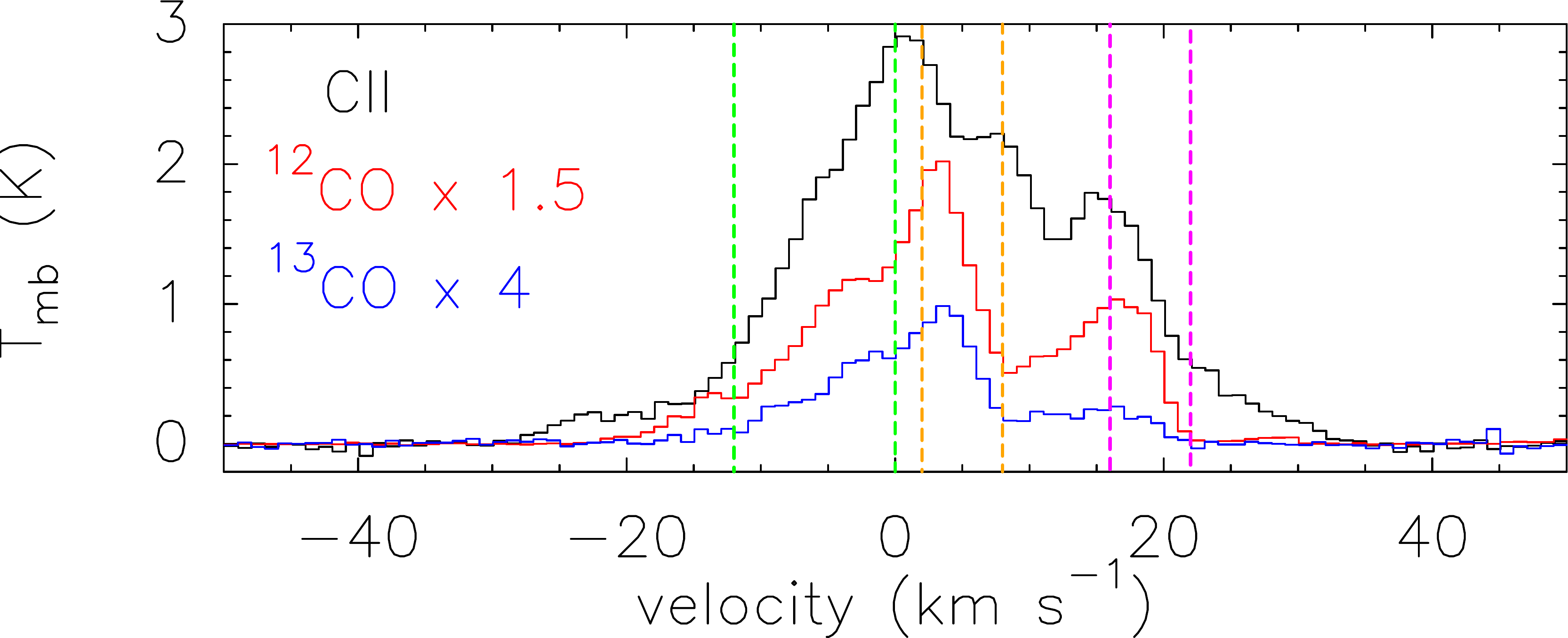}

\caption{Average spectra of \cii\,, $^{12}$CO and $^{13}$CO toward the whole mapped region of RCW~49. To highlight the structures seen in the velocity channel maps, we mark the boundaries of the shell's expansion (in green), the northern and southern clouds (in orange) and the ridge (in pink). The velocity integrated intensity maps of these regions are shown in Fig.~\ref{shell_nscl_ridge}.}  

\label{av_spectra} 
\end{figure}

\begin{figure*}[htp]
\centering
\includegraphics[width=45mm]{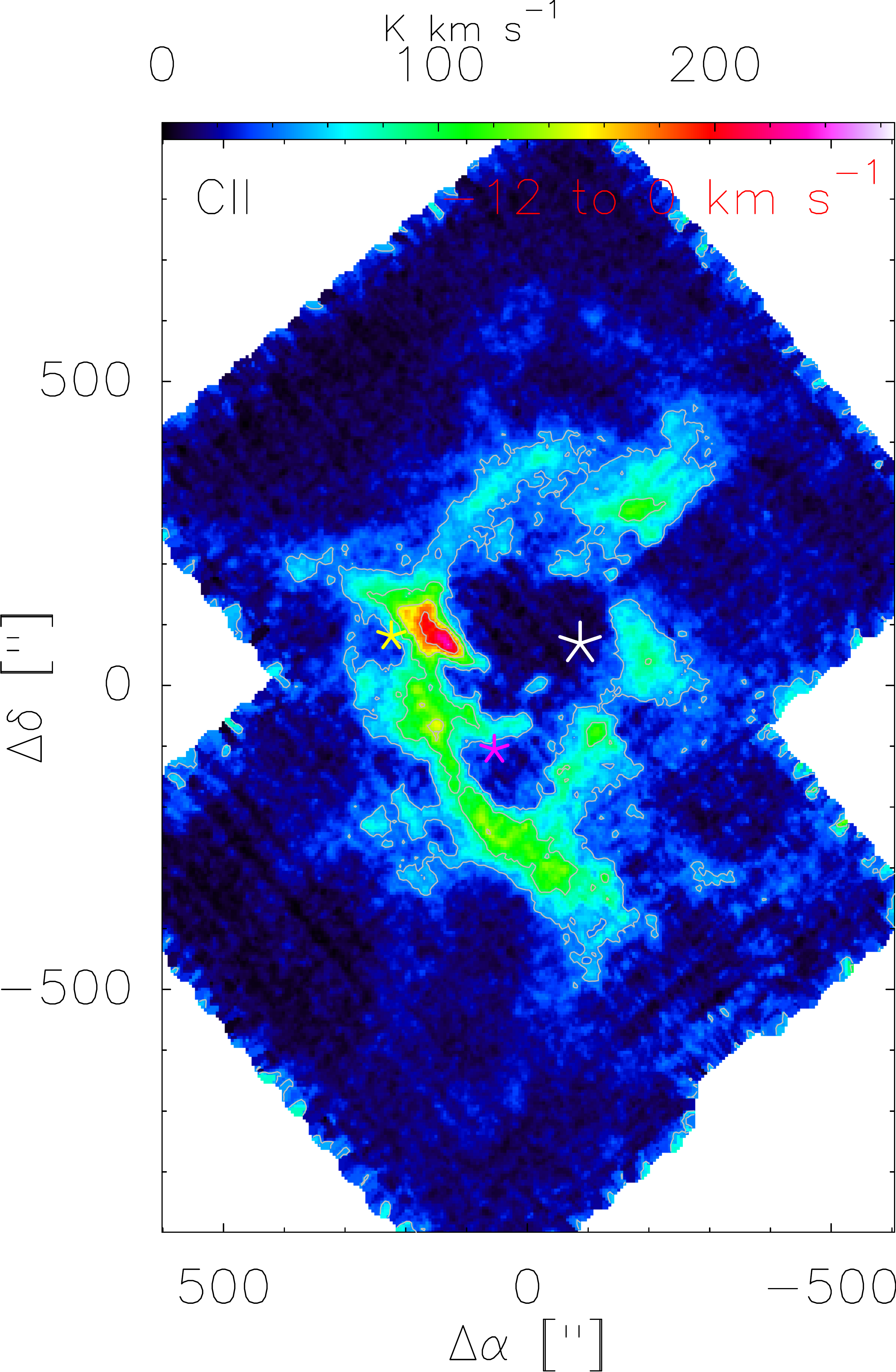}\quad
\includegraphics[width=45mm]{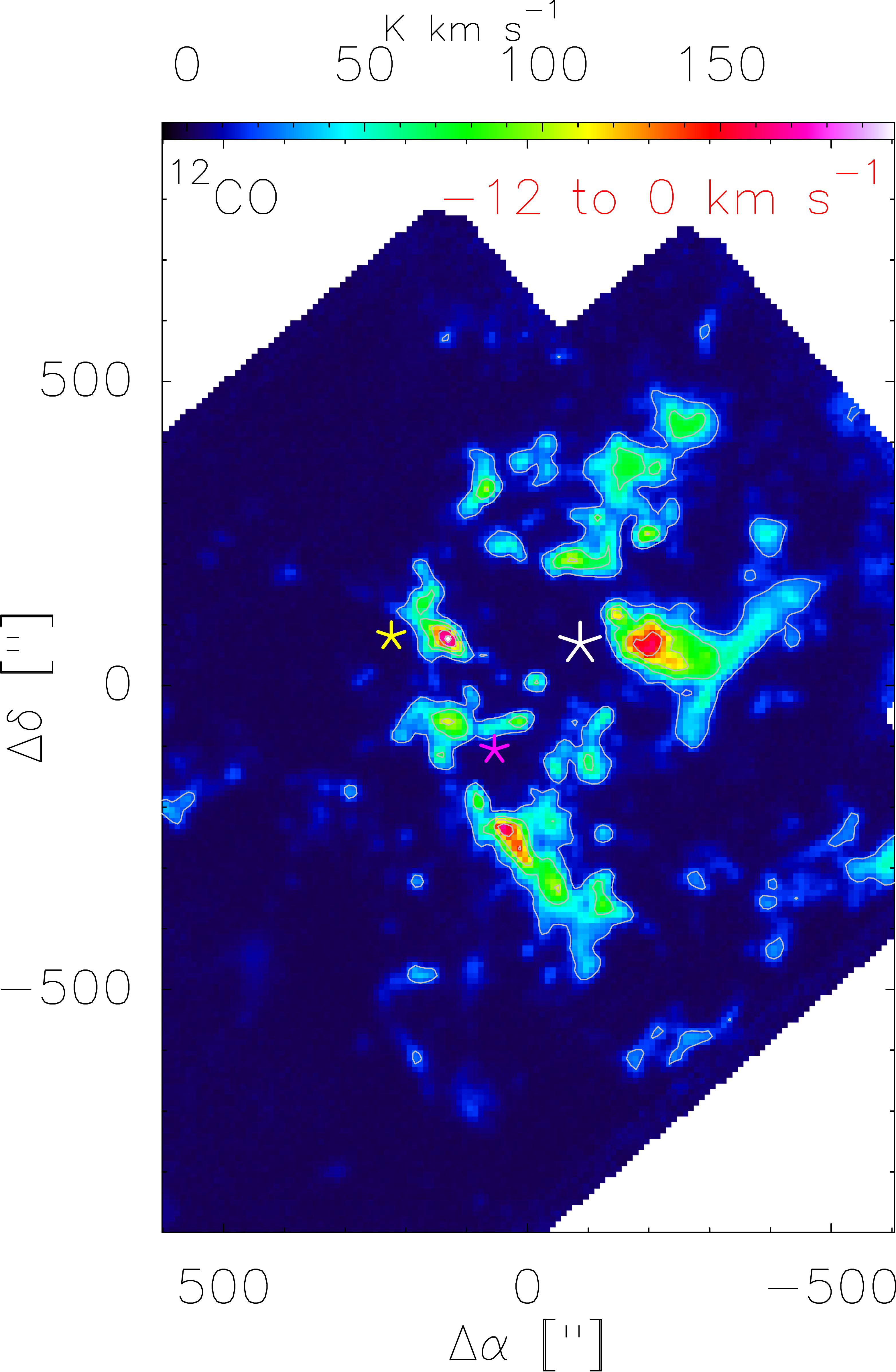}\quad
\includegraphics[width=45mm]{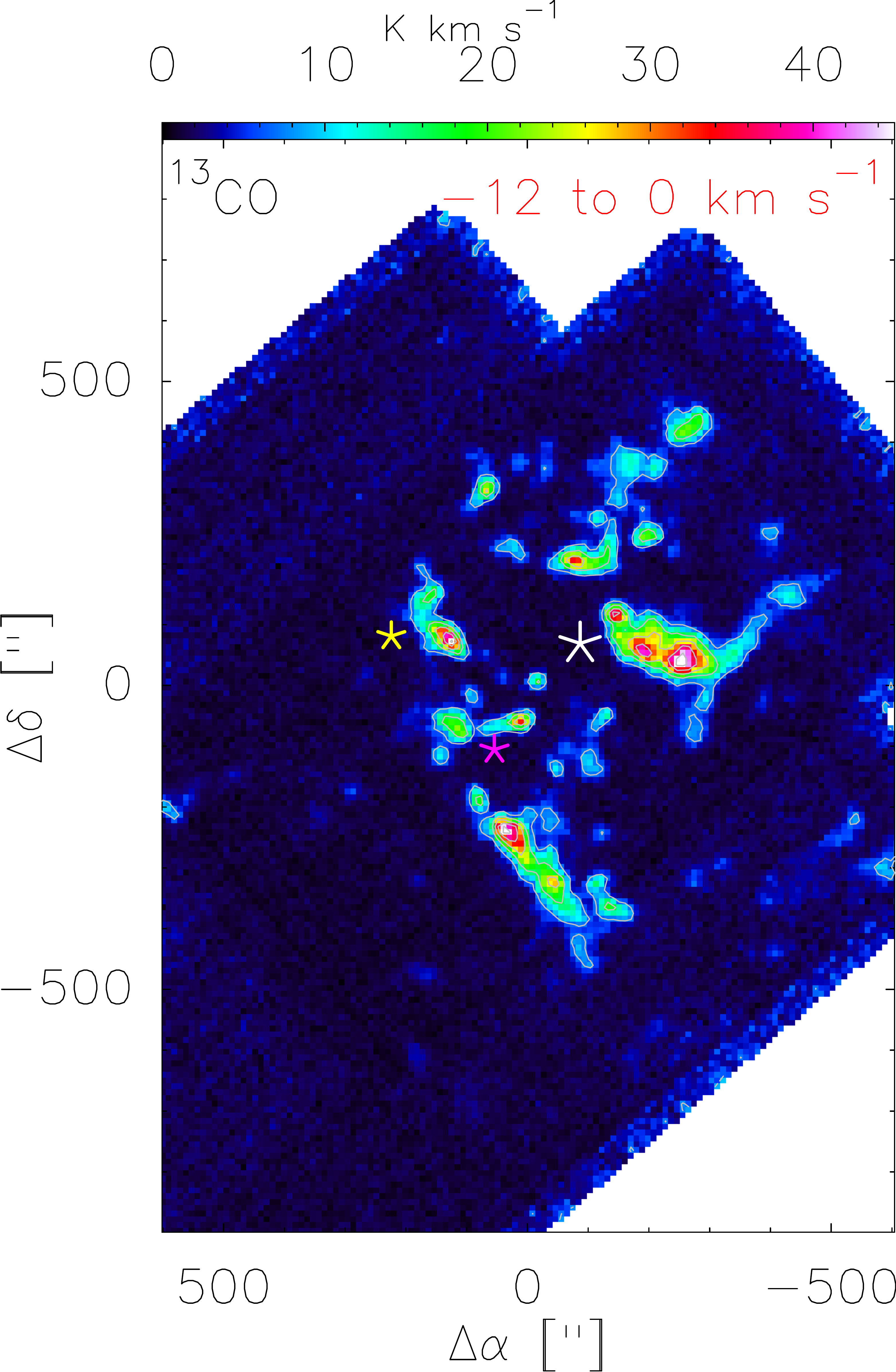}
\includegraphics[width=45mm]{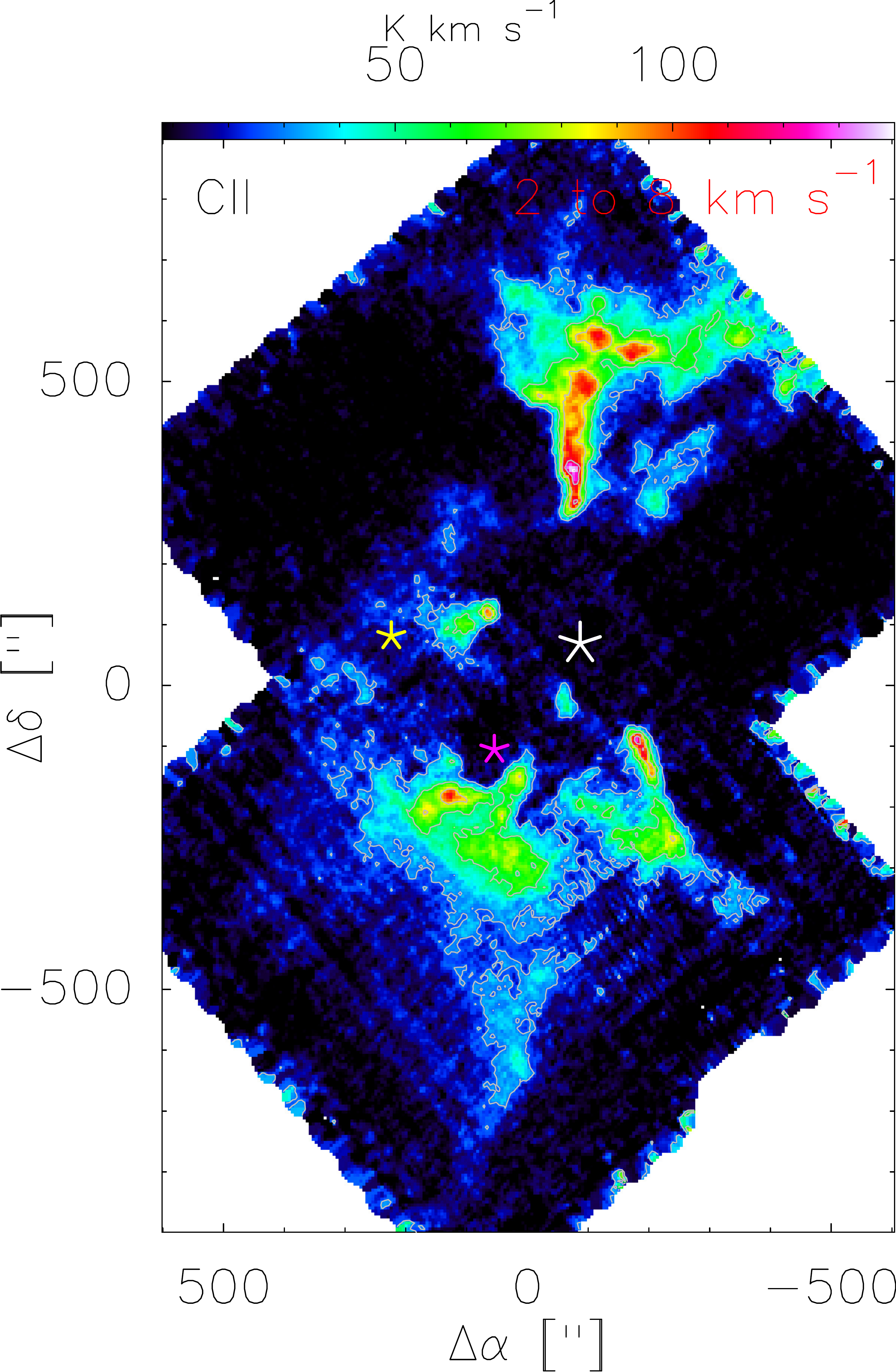}\quad
\includegraphics[width=45mm]{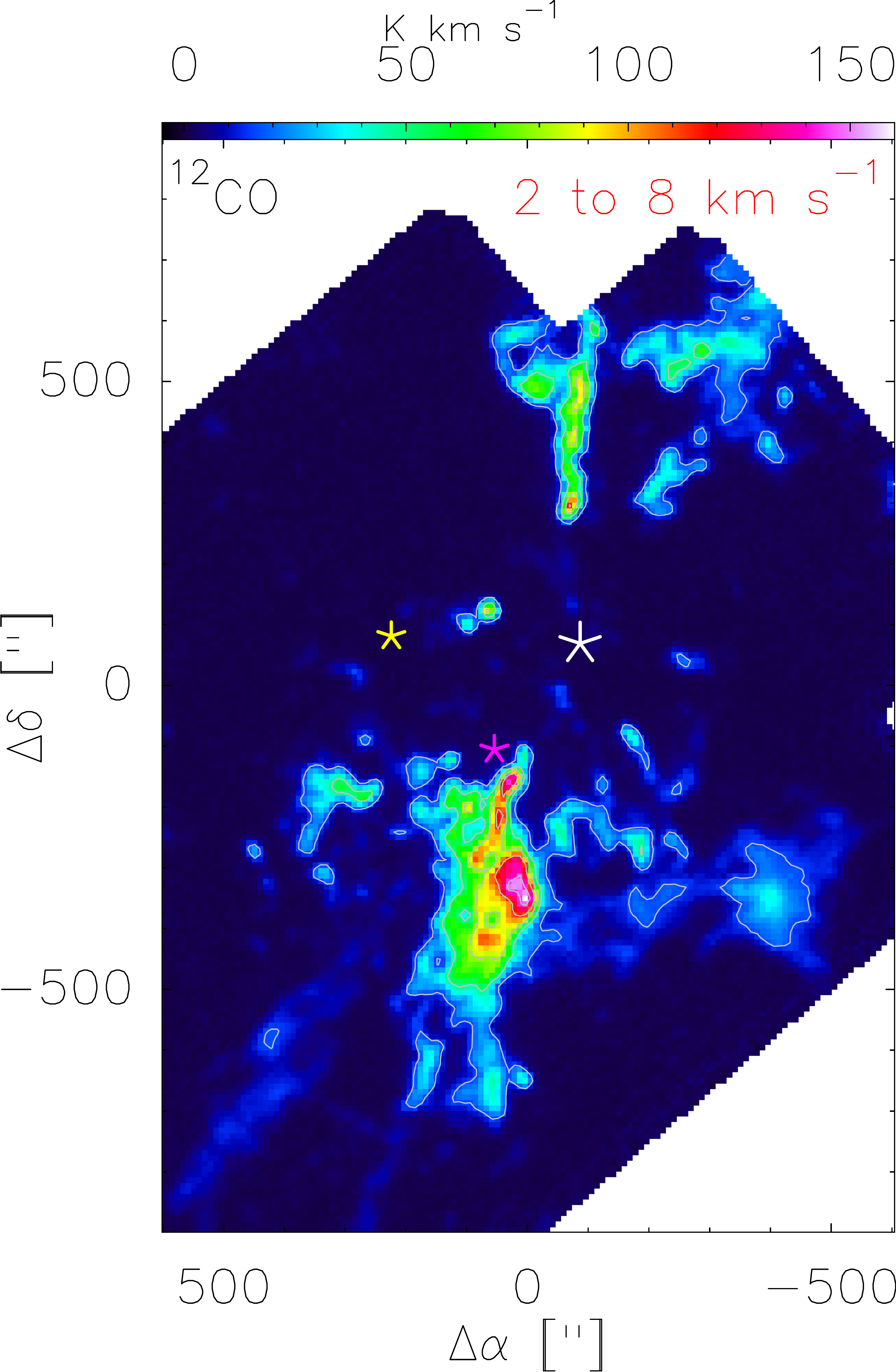}\quad
\includegraphics[width=45mm]{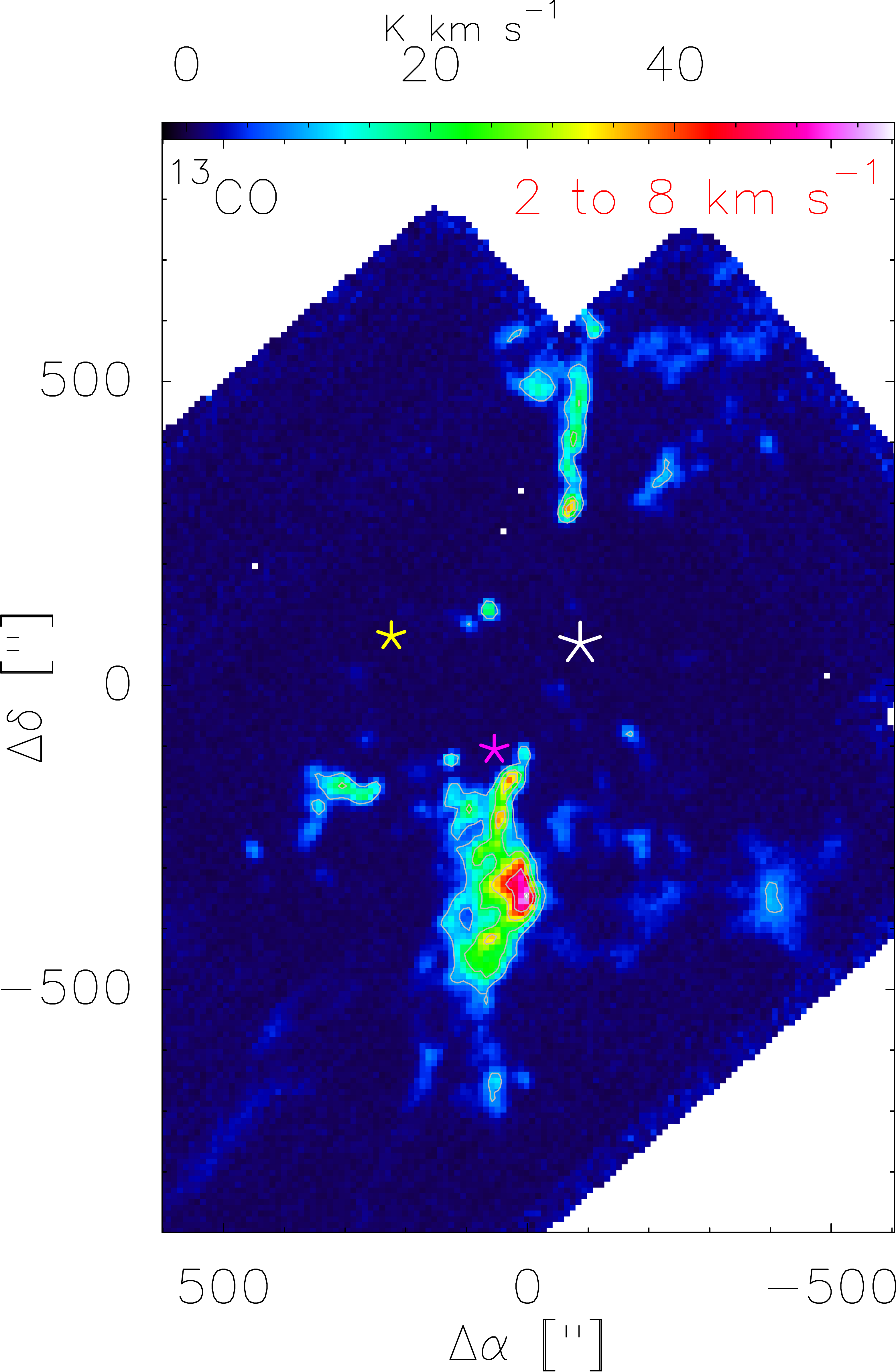}
\includegraphics[width=45mm]{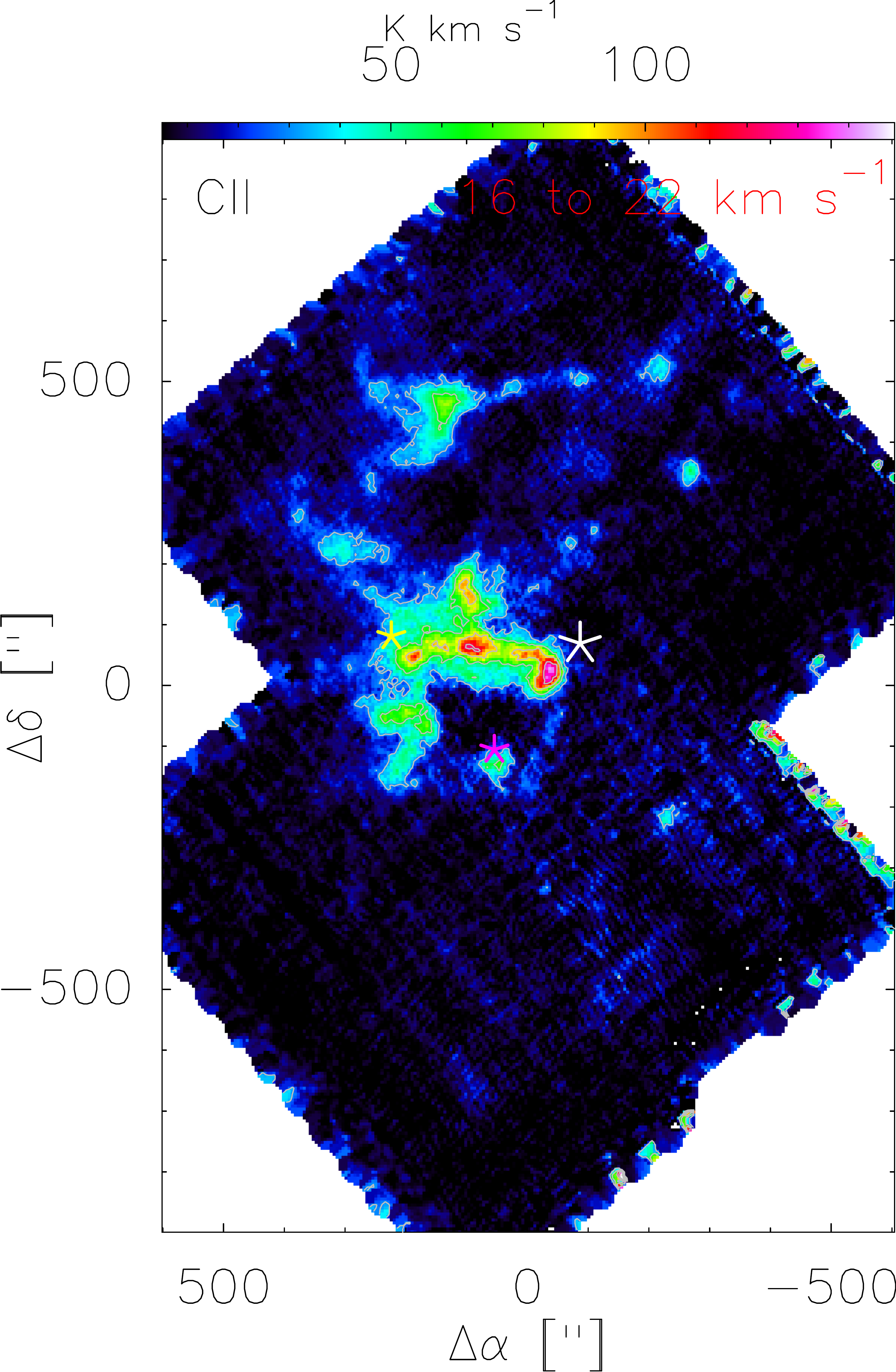}\quad
\includegraphics[width=45mm]{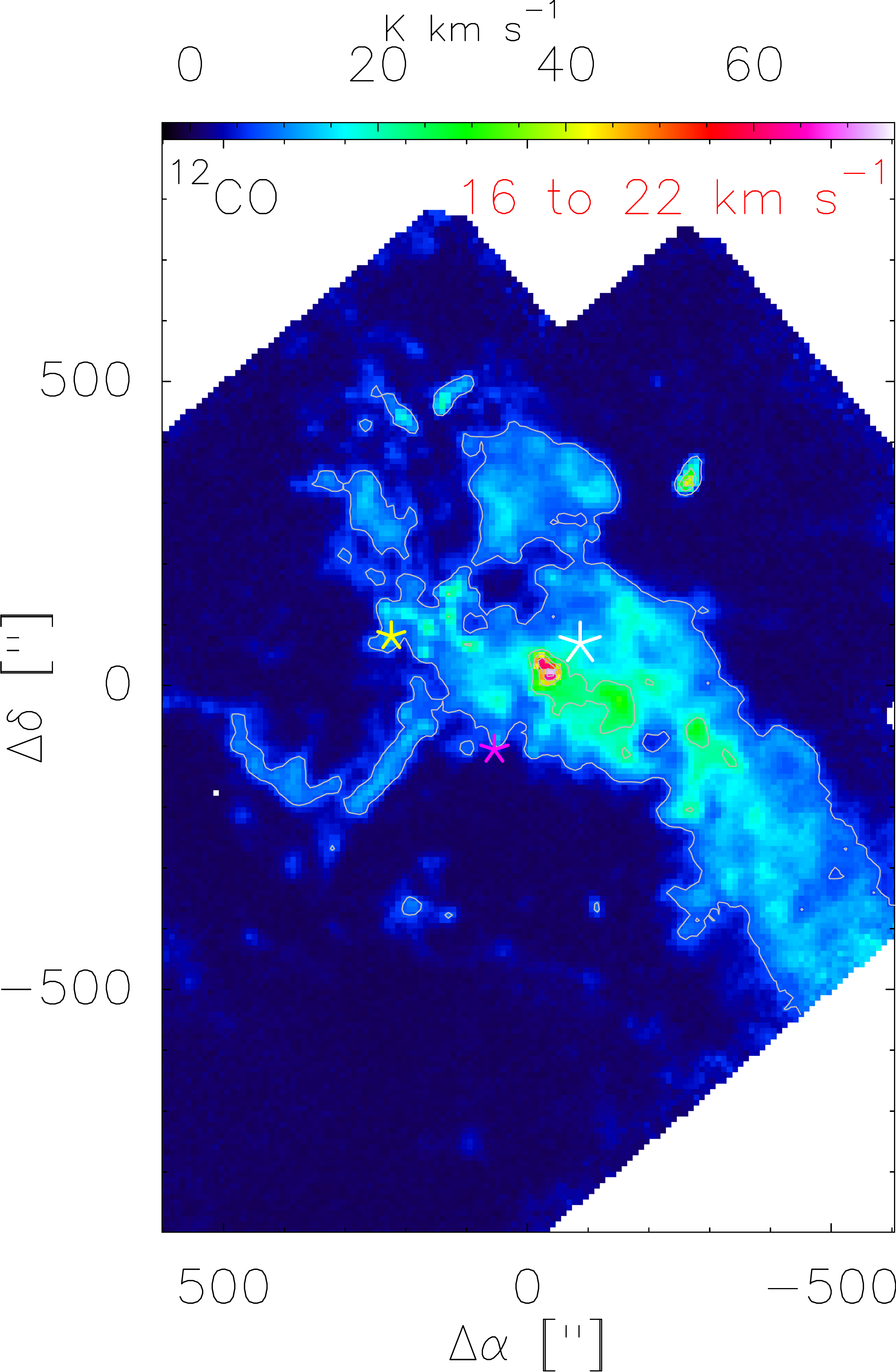}\quad
\includegraphics[width=45mm]{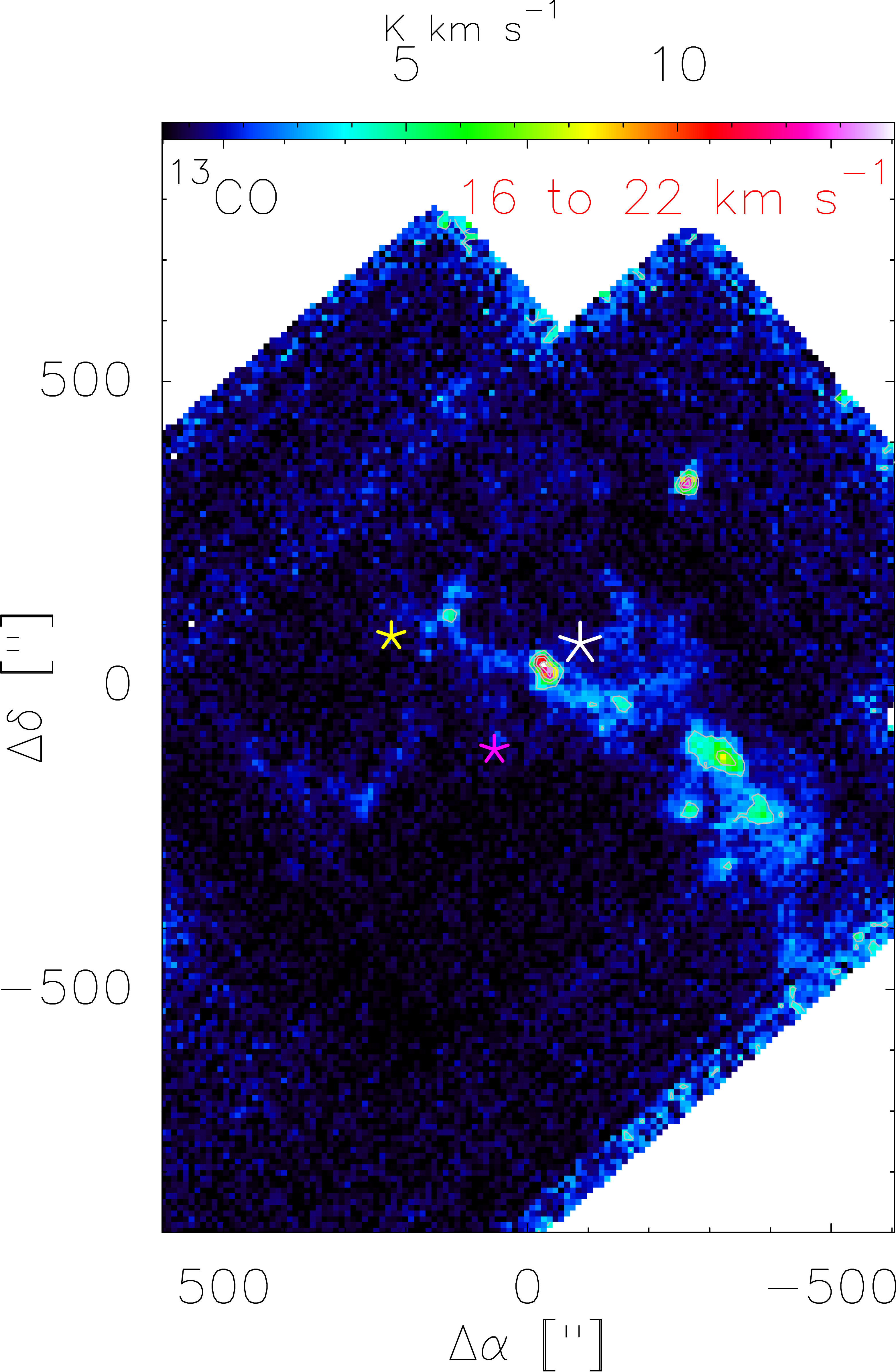}

\caption{Left to right: Velocity integrated intensity maps of \cii\,, $^{12}$CO and $^{13}$CO of the shell's expansion within -12 to 0~km~s$^{-1}$ (top row), the northern and southern clouds within 2 to 8~km~s$^{-1}$ (middle row) and the ridge within 16 to 22~km~s$^{-1}$ (bottom row). The Wd2 cluster's center, the OV5 and the WR20b stars are marked with white, yellow and pink asterisks respectively.}  

\label{shell_nscl_ridge} 
\end{figure*}

\subsection{Ancillary data}

We present the images of the most relevant ancillary data toward RCW~49 in the lower panel of Fig.~\ref{vel-int-maps}. We started with the Galactic Legacy Infrared Mid-Plane Survey Extraordinaire (GLIMPSE, \citealt{2003PASP..115..953B}) 8~$\mu$m data observed with the Spitzer Space Telescope. The 8~$\mu$m emission arises from the PDR surface of dense molecular clouds where large hydrocarbon molecules, the Polycyclic Aromatic Hydrocarbons (PAHs), are excited by strong UV radiation resulting in fluorescent IR emission \citep{2008ARA&A..46..289T}. Next, we used the 70~$\mu$m data from the Herschel Space Archive (HSA), obtained within the Hi-GAL Galactic plane survey \citep{higal2010} observed the with Photodetector Array Camera and Spectrometer (PACS, \citealt{2010A&A...518L...2P}) aboard the Herschel Space Observatory \citep{2010A&A...518L...1P}. The 70~$\mu$m emission traces the warm interstellar dust exposed to FUV radiation from the stars. Further, we obtained the 870~$\mu$m dust continuum data from APEX Telescope Large Area Survey of the Galaxy (ATLASGAL, \citealt{2009A&A...504..415S}), performed with the Large APEX BOlometer CAmera (LABOCA) instrument of the 12~m APEX telescope. The 870~$\mu$m traces the cold and dense clumps shielded from FUV radiation by large amounts of dust extinction. Hence, this map provides a good probe of the earliest star formation sites. Lastly, we used the archival data from the Chandra X-ray Observatory using its primary camera, the Advanced CCD Imaging Spectrometer (ACIS) \citep{2003SPIE.4851...28G}. Diffuse X-ray (0.5--7~keV) structures surrounding massive star-forming regions have been shown to trace hot plasmas from massive star feedback (e.g. \citealt{2019ApJS..244...28T}).    

\section{Results}

\subsection{Multi-wavelength overview of RCW~49}
In order to investigate the relation between the atomic and molecular components of the gas, we show the emission toward RCW~49 in different transitions
and continuum wavelengths in Fig.~\ref{vel-int-maps}.

A study of the large-scale $^{12}$CO (2 $\to$ 1) distribution by \citet{2009ApJ...696L.115F} identified two molecular clouds ($-$11 to 9 and 11 to 21~km~s$^{-1}$) in RCW~49 and suggested their collision ($\sim$ 4~Myr ago) to be responsible for triggering the formation of Wd2. The cloud within $-$11 to 9~km~s$^{-1}$ has a mass of (8.1 $\pm$ 3.7) $\times$ 10$^4$~\(\textup{M}_\odot\) and extends over a range of $\sim$ 26.1~pc in the north-south direction and $\sim$ 21.6~pc in the east-west direction. The cloud within 11 to 21~km~s$^{-1}$ has a mass of (9.1 $\pm$ 4.1) $\times$ 10$^4$~\(\textup{M}_\odot\) and extends over a range of $\sim$ 18.3~pc in the north-south direction and $\sim$ 21.6~pc in the east-west direction \citep{2009ApJ...696L.115F}. The $-$11 to 9~km~s$^{-1}$ cloud further includes two seemingly different clouds: $-$11 to 0~km~s$^{-1}$ and 1 to 9~km~s$^{-1}$ \citep[Fig~1~(c) and (d)]{2009ApJ...696L.115F}, but were analysed as one. This is perhaps due to the unavailability of high spatial resolution in the large scale $^{12}$CO data. We manually outline the 1 to 9~km~s$^{-1}$ and 11 to 21~km~s$^{-1}$ clouds in the right panel of Fig.~\ref{rgb-cii-8-870}. The $-$11 to 0~km~s$^{-1}$ cloud will be discussed in detail in the next sections of this paper. \citet{2004ApJS..154..322C} studied RCW~49 in mid-IR wavelengths to investigate its dust emission morphology and identified distinct regions as a function of the angular radius with respect to Wd2. It can be seen in Fig.~\ref{vel-int-maps} that all tracers, except the X-ray emission, are devoid of any emission in the immediate surrounding of Wd2. This emission free region is filled by hot plasma (temperature of $\sim$ 3 $\times$ 10$^6$~K and density of $\sim$ 0.7~cm$^{-3}$, see Sect.~3.5.2) as evident from its very bright X-ray emission (as seen in Fig.~\ref{vel-int-maps}). Moving away from Wd2, the 8 and 70~$\mu$m emission becomes brighter and marks the so called ``transition boundary'' (at $\sim$ 5~pc from Wd2, shown in Fig~\ref{rgb-cii-8-870}, right panel), a ring-like structure opened to the west \citep[Fig.~1]{2004ApJS..154..322C}. Dense cores are traced by ATLASGAL 870~$\mu$m. A dense ridge structure (pointed in Fig.~\ref{rgb-cii-8-870}, left panel), running from south to east of Wd2, is particularly prominent in the ATLASGAL emission map. This ridge is also visible in the 8~$\mu$m PAH and 70~$\mu$m dust continuum emission but not very prominent in the CO (3 $\to$ 2) emission maps. It is important to verify the spatial and spectral information given by $^{12}$CO observations with $^{13}$CO because in general, $^{12}$CO is optically thick and could be impacted by opacity effects causing self absorption, etc. Thus, the optically thin $^{13}$CO is used to confirm the results obtained from $^{12}$CO observations. In addition, the combination of $^{12}$CO and $^{13}$CO can be used to determine the molecular gas mass. In contrast, the dense cores to the west and north are also recognizable in CO (3 $\to$ 2) emission maps but not so prominent in 8~$\mu$m and 70~$\mu$m emission. Our $^{12}$CO (3 $\to$ 2) map follows the larger scale $^{12}$CO (2 $\to$ 1) emission distribution as reported in \citet{2009ApJ...696L.115F}.
In addition to the above mentioned bright emission, both have a slightly fainter emission toward the west of Wd2, which is absent in the ATLASGAL emission. 

The \cii\ emission map shows similarities to these structures. It reveals a lack of emission immediately surrounding Wd2. One of the most prominent structures in the \cii\ map is a bright and wide arc (labelled ``shell'' in Fig.~\ref{rgb-cii-8-870}, left panel) of emission running from the east to the south which is quite prominent in the channel maps from -14 to -6~km~s$^{-1}$ (see section 3.2). At higher velocities the \cii\ emission is more and more dominated by the ridge southeast from Wd2 and a separate, dense structure to the north. The brighter \cii\ peaks in these morphological structures have counterparts in the CO maps. We notice that the ridge structure is much less bright in the \cii\ line than in the 8, 70, and 870~$\mu$m maps. 
This can also be viewed in the red green blue (RGB) image of \cii\,, 8~$\mu$m and 870~$\mu$m emission shown in Fig.~\ref{rgb-cii-8-870}. The difference in the behavior of these structures in the various maps reveals the complex mechanical and radiative interaction of the Wd2 cluster with the surrounding molecular gas. In this paper, we will use the kinematic information in the \cii\ and CO emission maps to focus on the kinetics and energetics of the arc-like emission structure. In a future paper, we will examine the other emission components present in this data.\\

\subsection{Velocity channel maps}
A spatial comparison of the \cii\,, $^{12}$CO and $^{13}$CO emission in different velocity channels can be seen in Figs.~\ref{cii-chan}, \ref{co-chan} and \ref{13co-chan}. In the \cii\ channel maps (Fig.~\ref{cii-chan}), we can see that a shell like structure starts to develop from $\sim$ -24~km~s$^{-1}$ and it appears to be expanding in the velocity range of $\sim$ -12 to 0~km~s$^{-1}$. A red blue image is shown to depict this expansion in the Appendix~\ref{sec:shell_exp_appen}, Fig.~\ref{rb_-12_2}. The eastern arc of the shell is well defined compared to the western arc, which appears to be broken. In the velocity range of 2 to 12~km~s$^{-1}$, we see \cii\ emission confined to the north and south. The ridge as we discussed in Sect.~3.1 is seen in the velocity range of 16 to 22~km~s$^{-1}$. 

Channel maps of $^{12}$CO and $^{13}$CO (Figs.~\ref{co-chan} and \ref{13co-chan}) show similar velocity structures. The shell seen in the \cii\ channel maps is also outlined by fragmented emission from both $^{12}$CO and $^{13}$CO, starting from about -12 to 0~km~s$^{-1}$. Similar to the \cii\ emission, the eastern side of the shell is more apparent than its western counterpart. The northern and southern structures are spread out over a smaller velocity range (2 to 8~km~s$^{-1}$) compared to the \cii\ emission. Though the ridge is not very distinct in CO emission, it appears to be associated with the molecular cloud traced by $^{12}$CO for velocities greater than 16~km~s$^{-1}$.

\subsection{Different structures in RCW~49}
Figure~\ref{av_spectra} shows the average spectra of \cii\,, $^{12}$CO and $^{13}$CO toward all of the mapped regions shown in Fig~\ref{vel-int-maps}. We mark the boundaries of three main structures that we identify in the channel maps (Figs.~\ref{cii-chan}, \ref{co-chan} and \ref{13co-chan}): the shell's expansion, the northern and southern clouds and the ridge. Figure~\ref{shell_nscl_ridge} shows these structures as seen in velocity integrated intensity maps of \cii\,, $^{12}$CO and $^{13}$CO. The top row of Fig.~\ref{shell_nscl_ridge} shows the integrated intensity of the velocity range within which the expansion of the shell is clearly visible (in Fig.~\ref{cii-chan}) and we discuss this shell in detail in the further sections. The northern and southern clouds and the ridge are shown in the middle and bottom rows. These two structures are similar to the clouds as reported by \citet{2009ApJ...696L.115F} and as shown in Fig.~\ref{rgb-cii-8-870}. 

\subsection{Spectra toward different offsets}
\begin{figure*}[htp]
\includegraphics[width=100mm]{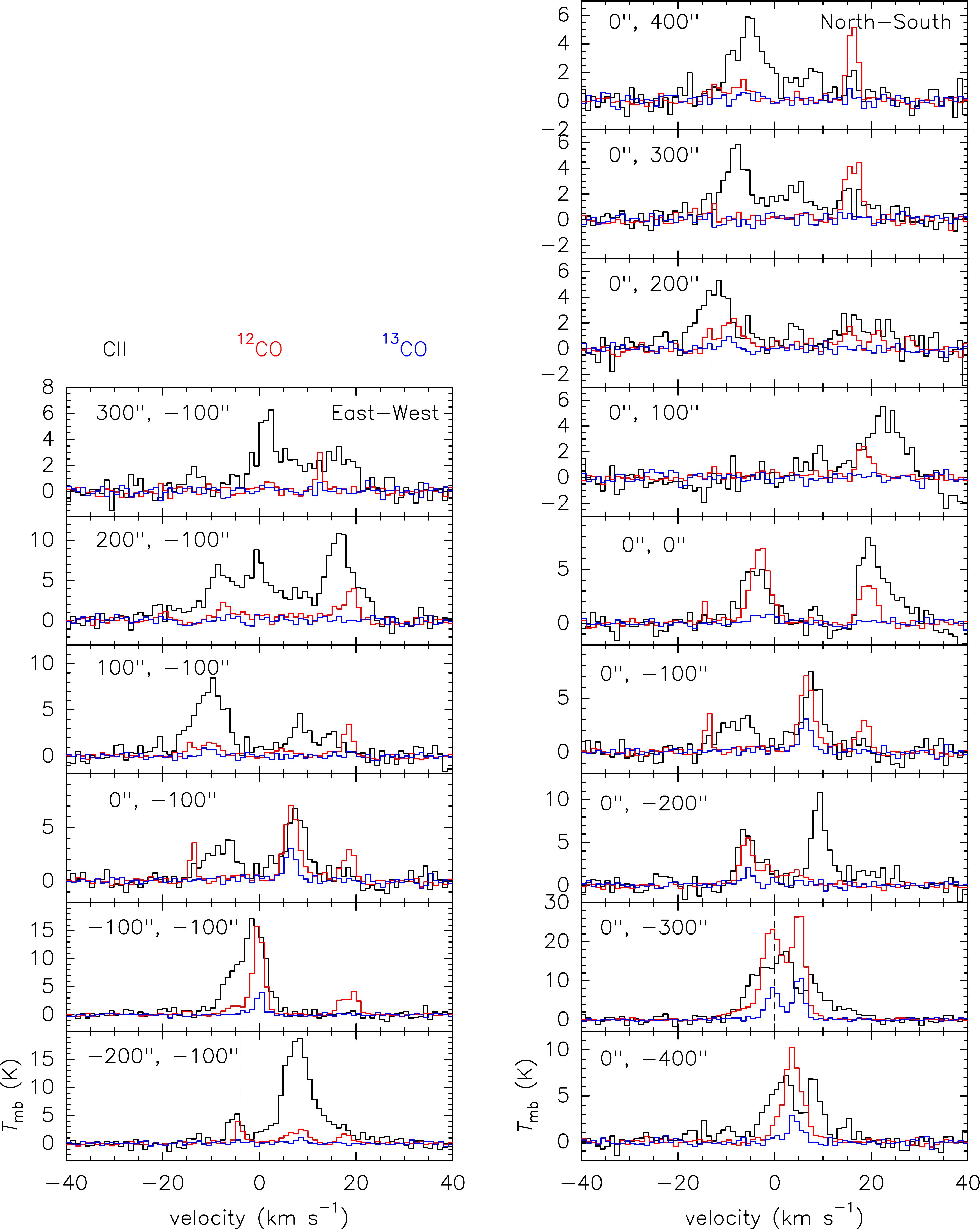}\qquad
\includegraphics[width=73mm]{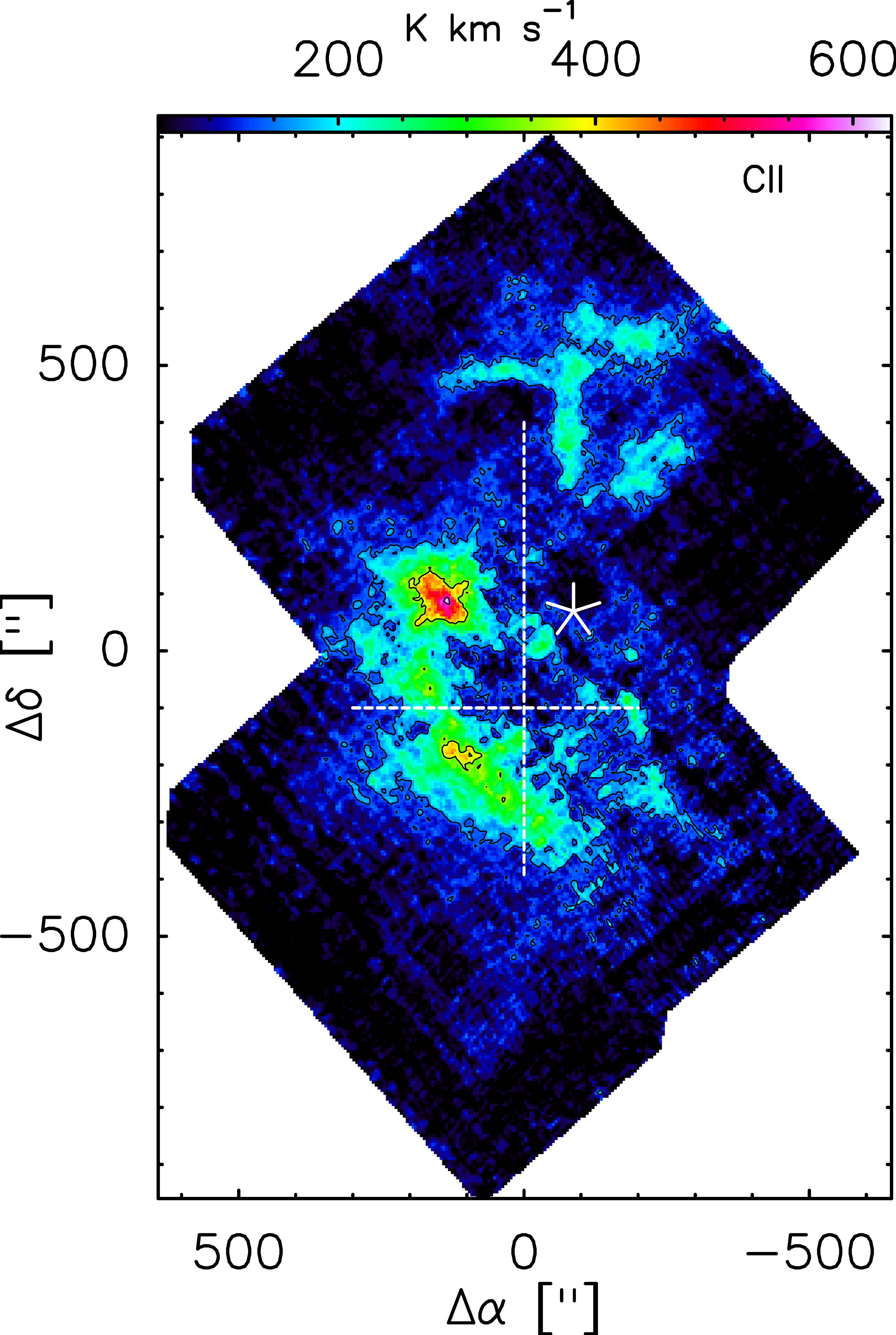}

\caption{Spectra (left and middle panels) of \cii\,, $^{12}$CO and $^{13}$CO toward different offsets along the horizontal and vertical lines shown on the \cii\ velocity (-25 to 30~km~s$^{-1}$) integrated intensity map in the right panel. The presented spectra are smoothed to a velocity resolution of 1~km~s$^{-1}$. The sequential shift in the peak of the blue-shifted component is also marked with vertical dashed lines in some panels. The blue-ward shift of the expanding shell is quite apparent when comparing these spectra (see Sections 3.2, 3.3 and 3.4).}  

\label{spectra} 
\end{figure*}

Figure~\ref{spectra} (left and middle panels) shows examples of observed spectra of \cii\,, $^{12}$CO and $^{13}$CO toward different offsets along the horizontal and vertical cuts as shown in the right panel. These cuts were chosen to visualize the spectral line profiles of the observed species toward the shell, which we see in the velocity channel maps (in Figs.~\ref{cii-chan}, \ref{co-chan} and \ref{13co-chan}).

There are multiple velocity components toward any position but by examining closely the line profiles of the blue-shifted velocity component, we can follow the shell's progression. As we move along the horizontal cut (at $\Delta\delta$ = $-$100$\arcsec$) shown in the right panel of Fig.~\ref{spectra}, we see that the blue-shifted velocity component of the \cii\ line sequentially starts to shift its peak from $\sim$ 0~km~s$^{-1}$ (at $\Delta\alpha$ = 300$\arcsec$) to $\sim$ $-$11~km~s$^{-1}$ (at $\Delta\alpha$ = 100$\arcsec$) and back to $\sim$ $-$4~km~s$^{-1}$ (at $\Delta\alpha$ = $-$200$\arcsec$). Thus, tracing the shell's expansion. For the $^{12}$CO emission, the blue-shifted velocity component roughly follows the \cii\ line profiles. It starts to show up at $\Delta\alpha$ = 200$\arcsec$, sequentially shifting its peak to $\sim$ 14~km~s$^{-1}$ (at $\Delta\alpha$ = $-$100$\arcsec$) and back to $\sim$ $-$4~km~s$^{-1}$ (at $\Delta\alpha$ = $-$200$\arcsec$), similar to \cii\ line. As mentioned earlier (in Sect.~3.1), $^{12}$CO is usually optically thick, thus, we need to examine $^{13}$CO spectral line profiles to confirm the presence of different velocity components seen in $^{12}$CO spectra. The blue-shifted velocity component of $^{13}$CO emission line intensity is low and is not detected with our S/N. However, it can be seen that toward the lines-of-sight where the red-shifted velocity component of $^{12}$CO is relatively brighter, $^{13}$CO is detectable and follows $^{12}$CO's line profile. Thus, we expect $^{13}$CO emission line to have similar profiles for its blue-shifted velocity component as well.


A similar trend is observed in the spectra reported along the vertical cut (at $\Delta\alpha$ = 100$\arcsec$). A shift in the peak of the blue-shifted velocity component of \cii\ can be seen starting from $\sim$ 0~km~s$^{-1}$ (at $\Delta\delta$ = $-$300$\arcsec$) to $\sim$ $-$13~km~s$^{-1}$ (at $\Delta\delta$ = 200$\arcsec$) and then back to $\sim$ $-$5~km~s$^{-1}$ (at $\Delta\delta$ = 400$\arcsec$). At $\Delta\delta$ = 100$\arcsec$, we see no emission from the shell ($<$ 0 km~s$^{-1}$) as evident from both the spectrum and its spatial distribution described in Sect.~3.5.
The $^{12}$CO and $^{13}$CO emission also follow a similar trend as the spectra along the horizontal cut. 

\begin{figure*}[htp]
\centering
\includegraphics[width=150mm]{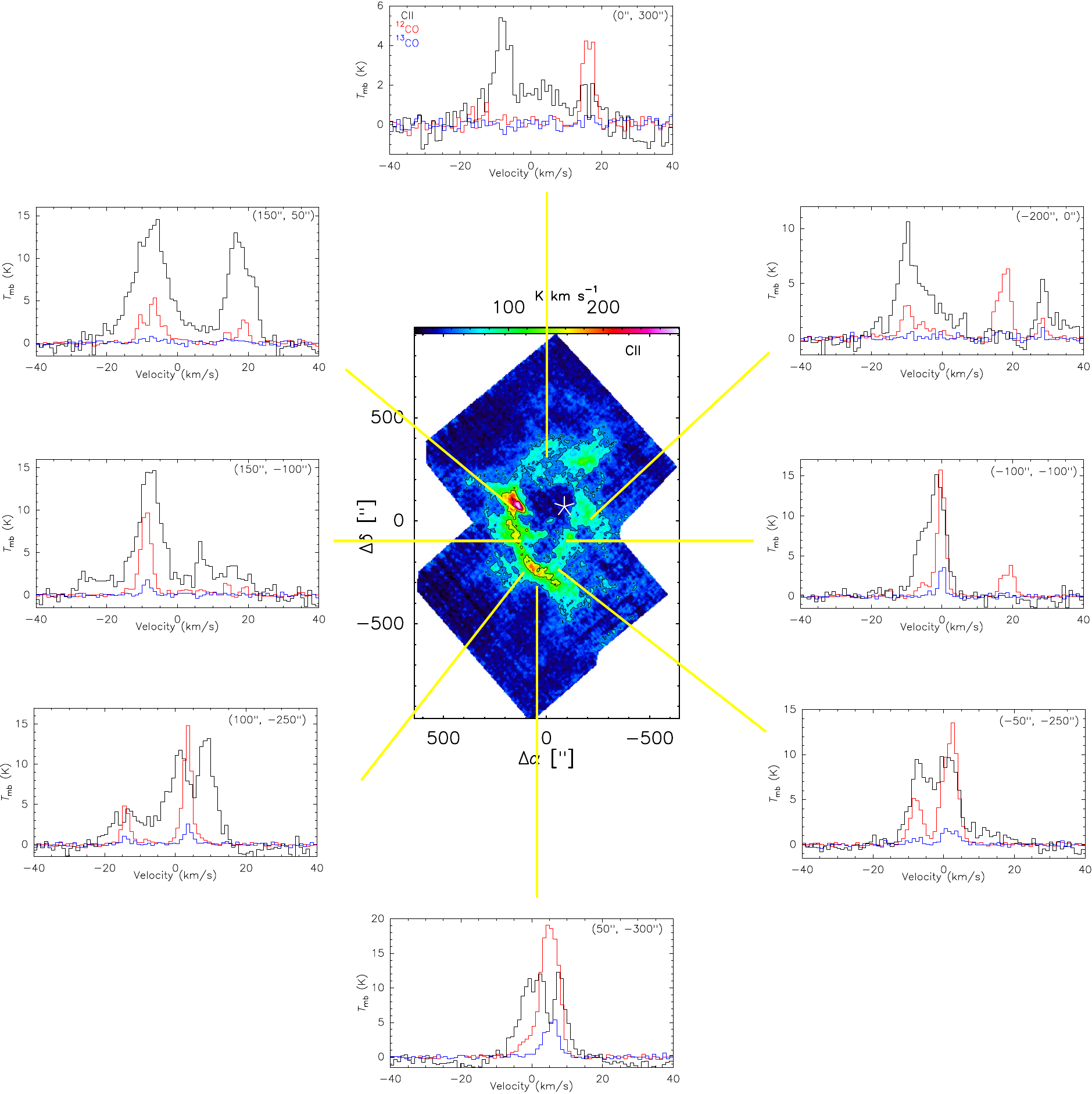}

\caption{Spectra of \cii\, (black), $^{12}$CO (red) and $^{13}$CO (blue) toward different offsets along the shell, shown around the velocity integrated (from -25 to 0~km~s$^{-1}$) intensity map of \cii\,.}  

\label{rolf-spec} 
\end{figure*}

In addition to the spectra shown in Fig.~\ref{spectra}, we selected a few more positions specifically along the shell to examine the spectral line profiles of the observed species. Figure~\ref{rolf-spec} shows the spectra of \cii\,, $^{12}$CO and $^{13}$CO at different offsets marked on the shell as seen from the velocity (-25 to 0~km~s$^{-1}$) integrated intensity map of \cii\,. Most of the spectra show a prominent blue-shifted ($<$ 0~km~s$^{-1}$) velocity component, which is tracing the shell.    

In contrast to the blue-shifted velocity component, the red-shifted velocity component ($>$ 15~km~s$^{-1}$) of the observed spectra does not follow any obvious trend. It corresponds to the ridge as discussed in Sect.~3.3 and shown in Fig.~\ref{shell_nscl_ridge} bottom row. The velocity components of \cii\,, $^{12}$CO and $^{13}$CO that lie within 2 to 12~km~s$^{-1}$ are part of the northern and southern clouds (discussed in Sect.~3.3 and shown in Fig.~\ref{shell_nscl_ridge} middle panel). The spectra shown in Fig.~\ref{rolf-spec} at $\Delta\alpha$ = 50$''$, $\Delta\delta$ = -300$''$ show $^{12}$CO and $^{13}$CO line profiles peaking between the two peaks of \cii\ line, which is indicative of self-absorption at velocities $>$ 0~km~s$^{-1}$. This spectral behaviour could also be seen at an offset of $\Delta\alpha$ = 100$''$, $\Delta\delta$ = -250$''$. This suggests that the northern and southern clouds probably lie in front of the gas constituting the shell. 

\subsection{Expanding shell of RCW~49} \label{sec:shell_description}

\begin{figure*}[htp]
\centering
\includegraphics[width=180mm]{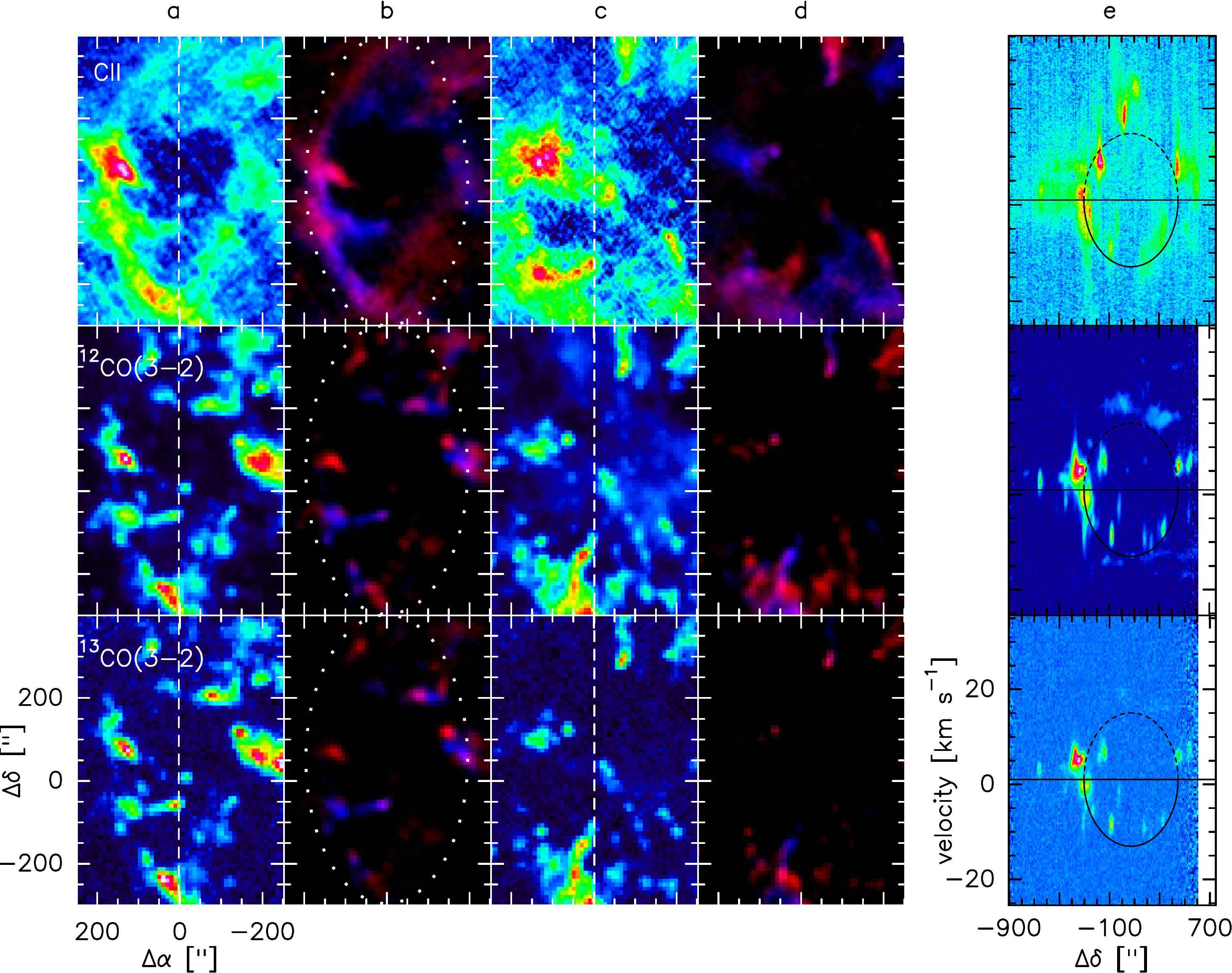}

\caption{The top, middle and bottom rows are for \cii\,, $^{12}$CO and $^{13}$CO, respectively. The maps shown here are smaller cutouts of the ones shown in Fig.~\ref{vel-int-maps}. Columns~a and c show the velocity integrated intensity maps of the observed species in the velocity ranges of $-$25 to 0~km~s$^{-1}$ and 0 to 30~km~s$^{-1}$, respectively. Column~b shows red blue (RB) velocity channel maps of \cii\,, $^{12}$CO and $^{13}$CO ranges from $-$12 to $-8$~km~s$^{-1}$ (blue) and from $-$8 to $-4$~km~s$^{-1}$ (red), respectively. The shell is marked with white dots in the three rows of column b. Similarly, column~d shows RB velocity channel maps of \cii\,, $^{12}$CO and $^{13}$CO from 0 to 4~km~s$^{-1}$ (blue) and from 4 to 8~km~s$^{-1}$ (red), respectively. Column~e shows the pv diagrams along the vertical white dashed line cuts in the columns~a and c. The predicted ellipse is shown on the pv diagrams for the blue-shifted part (in solid curve) and it's flipped (in dashed curve) to the red-shifted velocity structures.}


\label{bub-exp} 
\end{figure*}

Owing to the high spectral and spatial resolution of our \cii\,, $^{12}$CO and $^{13}$CO data, we were able to disentangle different components of gas along the same line-of-sight. The \cii\ channel maps (Fig.~\ref{cii-chan}) and the spectra displayed in Fig.~\ref{spectra} reveal a blue-ward expanding shell to the east of Wd2. The two (in red and blue) \cii\ velocity channel maps displayed in Fig.~\ref{bub-exp}~b outline this arc structure particularly well. The west side somewhat mirrors this arc-like structure but it is clearly a less coherent structure. The blue-shifted channel maps of the $^{12}$CO and $^{13}$CO emission reveal a highly fragmented clumpy distribution coincident with this limb-brightened shell. The position-velocity (pv) diagrams (Figs.~\ref{bub-exp}~e, \ref{fig:shell_pv} and \ref{pv-ver}) also reveal clearly the eastern arc as well as the fragmented western arc. The red-shifted (velocities $>$ 1~km~s$^{-1}$) counterpart of the east and west arcs are not very prominent in the channel maps (Figs.~\ref{cii-chan}, \ref{co-chan}, \ref{13co-chan} and 
\ref{bub-exp}~d), but there is evidence for a large scale, though highly fragmented, arc-like structure in the red-shifted velocities pv diagrams as well. Perusing the pv diagrams (Fig.~\ref{bub-exp}~e), we discern a spheroidal shell structure as outlined by large dots in Fig.~\ref{bub-exp}~b. Assuming that the shell is expanding isotropically, its projection on position-velocity space will be an ellipse. The maximum observed velocity and radius of the shell depend on the cosine of the angle between the center of the shell and a given cut along which a pv diagram is considered (for details see \citealt{2018PhDT.......111B}). The horizontal straight dashed line in the pv diagrams, represents the systemic velocity $\sim$ 1~km~s$^{-1}$, estimated by examining the velocity profiles of \cii\ emission toward various positions. The systemic velocity can vary by 1--2~km~s$^{-1}$ when looking at the spectra along different offsets. This is also consistent with the $^{12}$CO data. The maximum observed velocity of the shell is $\sim$ 13-14~km~s$^{-1}$, estimated from the observed spectra along different horizontal and vertical cuts through the shell. The center of the expanding shell is estimated from the intersection of the longest horizontal and vertical cuts at $\sim$ $\Delta\alpha$ = 0$\arcsec$, $\Delta\delta$ = 50$\arcsec$, which is $\sim$ 100$\arcsec$ east of Wd2. The resulting predicted ellipse is shown (solid curve in Fig.~\ref{bub-exp}~e) for our observed shell.
Comparing the horizontal and vertical cuts and their corresponding pv diagrams (in Fig.~\ref{pv-ver}), we get a vertical (north-south) radius of $\sim$ 7.5~pc and a horizontal (east-west) radius of $\sim$ 4~pc, such that the geometric mean elliptical radius will be $\sim$ 5.5~pc
and a thickness of $\sim$ 1~pc are estimated. Thus, the shell has expanded more in the north-south direction than the east-west direction.

In column~e of Fig.~\ref{bub-exp}, we flipped the ellipse (shown in dashed curves) of our blue-shifted shell and found that the actual structures at these higher velocities are not traced well by the ellipses. Instead the red-shifted part of the gas comprises dense clumps that are moving at higher speeds $\sim$ 20~km~s$^{-1}$. The red-shifted gas component, which we link to the northern and southern cloud structures and the ridge, will be discussed in a future paper.

\subsection{Dynamics of the shell}

\subsubsection{Mass of the shell} \label{sec:mass}

We have used several independent methods to estimate the mass of the expanding shell. Each of these methods relies on calculating the mass of the thin, limb-brightened arc of emission as identified on the \cii\ channel maps. We then use a geometric model to estimate the total mass of the shell. \\

\begin{figure*}[htp]
    \centering
    \includegraphics[width=\textwidth]{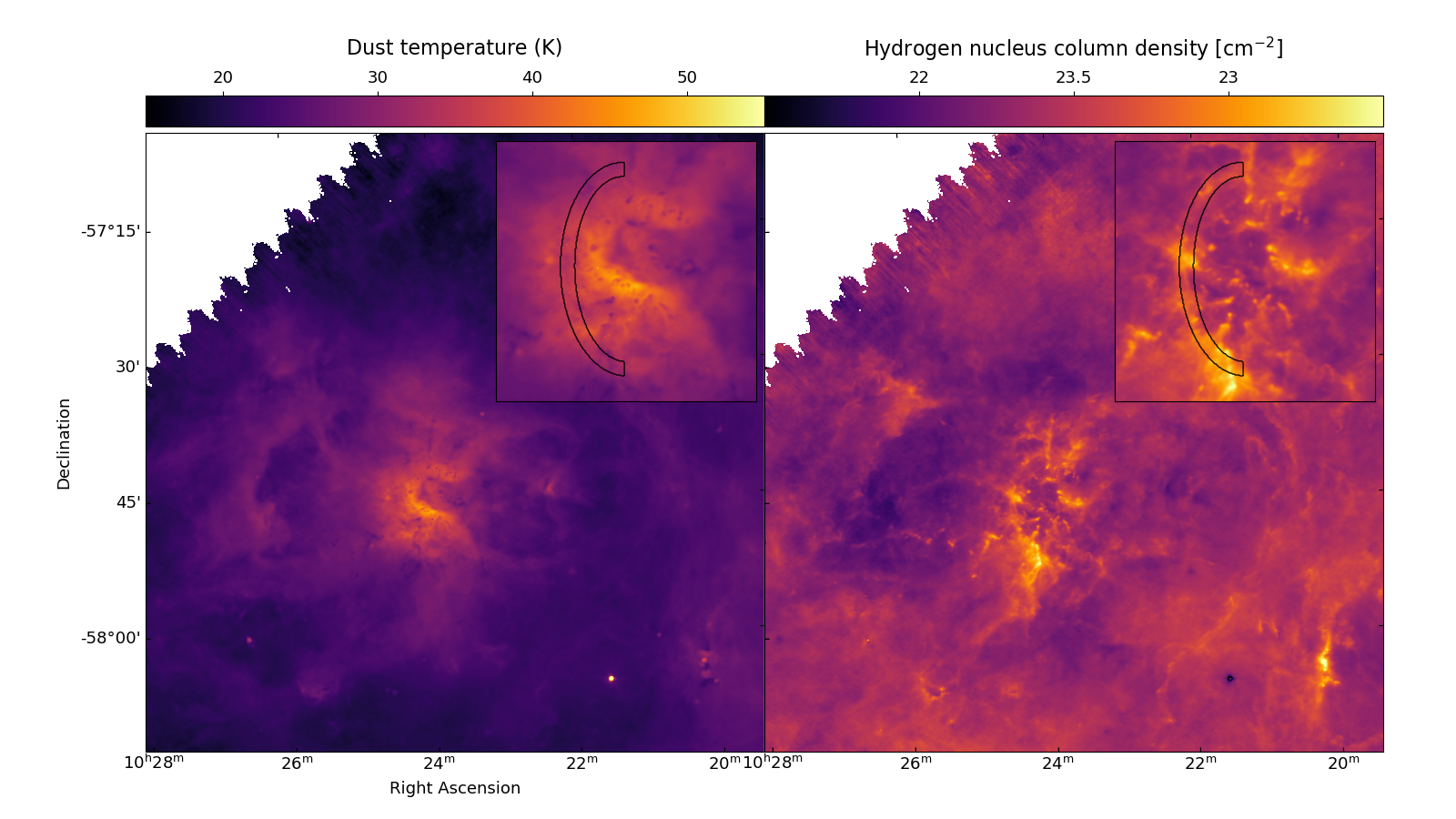}
    \caption{Map of temperatures \textit{(left)} and log-10 column densities \textit{(right)} derived pixel-by-pixel from the \herschel\ 70 and 160~$\mu$m data as described in Section~\ref{sec:mass}. The inset images in the upper right corners zoom in on the central region and show the half-ellipse mask used to estimate the shell mass from dust, \cii, and CO(3$-$2).}
    \label{fig:dust_mask}
\end{figure*}

In the first method, we used 70 and 160~\um\ data from \herschel\ PACS to estimate gas column densities via the dust column.
Illuminated dust throughout high-mass star forming regions is heated by the FUV radiation and re-emits this energy in the far infrared (FIR) \citep{1999RvMP...71..173H}.

The thermal FIR emission spectrum can be modeled as a modified black-body spectrum with spectral index $\beta$ in order to parametrize the emission in terms of dust (effective) temperature and optical depth.
With predictions from dust grain models, such as those of \citet{Draine2003a}, the derived optical depth can be converted to a hydrogen nucleus column density.
We are ultimately interested in deriving the mass in the shell, which is lined with PDRs and thus contains warmer dust.
The 70 and 160~$\mu$m PACS bands are more sensitive to warmer ($T > 20$~K) dust than the longer-wavelength SPIRE bands (250, 350, and 500~\um), so we elect to use only the 70 and 160~\um\ bands in our dust emission spectrum analysis.
\citet{Castellanos2014}, in their Section~4.3, make a comparable analysis of these FIR observations of RCW~49 using these two PACS bands.

Following the general technique of \citet{Lombardi2014}, we zero-point calibrated the PACS images by predicting the intensity in the PACS 70 and 160~\um\ bands at 5$'$ resolution using the \planck\ GNILC foreground dust model \citep{PlanckGNILC}.
We compare the \planck-predicted emission to the observed \herschel\ emission in order to determine the image-wide correction for each band.
Because we are making a significant extrapolation in predicting the shorter-wavelength PACS intensities using the \planck\ dust model, which is based on longer-wavelength \planck\ observations, we make this comparison under a mask excluding the warm central region.
This exclusion limits the comparison to lines of sight with low temperature variation compared to the central region and still includes a large area of $\sim 25$~K dust (see Figure~\ref{fig:dust_mask}) which is reasonably bright from the \planck\ wavelengths up to PACS 70~\um, maintaining the validity of our extrapolation.
A Gaussian curve is fitted to the distribution of differences between the predicted and observed intensity for each of the two PACS bands, and the fitted mean is assigned as the required zero-point correction.
The correction, a single number for each band, is added to each image.
The 70 and 160~\um\ intensities along the limb-brightened shell are of the order $\sim$ 0.5 to 2 $\times$ 10$^{4}$~MJy~sr$^{-1}$, so our applied zero-point corrections of 80 and 370~MJy~sr$^{-1}$, respectively, are no more than 10$\%$ of the total intensity in either band.

After zero-point correcting the two PACS images, we derive dust effective temperature and optical depth. We can model the emission with a modified black-body spectrum using Equations 1--3 from \citet{Lombardi2014}. In order to compare model emission with the PACS intensity measurements, we use Equation~8 from \citet{Lombardi2014} to integrate the model spectrum over the relative spectral response functions available for both the 70 and 160~\um\ PACS bands. In the present work, all source intensities and response functions which we discuss refer to the extended emission versions (as opposed to point source emission) where applicable. The PACS photometry is expressed as intensity (MJy/sr), already accounting for beam areas.
Following their Equation~8, we can write the mean intensity $\overline{I}_{i}$ measured in band $i$ as

\begin{equation} \label{eq:dust_bandpass}
    \overline{I}_{i} = \frac{\int I_{\nu}\, R_{\nu}\, \text{d}\nu}{\int (\nu_{i}/\nu)\, R_{\nu}\, \text{d}\nu }.
\end{equation}

Expressing the above function of $I_{\nu}$ as a ``bandpass function'' $BP_{i}$ of the incident source intensity $I_{\nu}$ and combining with their Equation~1, we rewrite $\overline{I}_{i}$ as
 
\begin{equation} \label{eq:dust_intensity}
 \overline{I}_{i} = BP_{i}\Big[ I_{\nu} \Big] = BP_{i}\Big[ B_{\nu}(T) (1 - e^{-\tau_{\nu}}) \Big]    
\end{equation}
 
where $B_{\nu}(T)$ is the Planck function in Equation~2 by \citet{Lombardi2014} and $\tau_{\nu}$ is modeled as a power law with spectral index $\beta$, as given in their Equation~3.

\begin{equation} \label{eq:dust_taudefinition}
\tau_{\nu} = \tau_{0} \big(\nu / \nu_{0} \big)^{\beta}.    
\end{equation}

We adopt $\nu_{0} = 1874$~GHz, corresponding to 160~\um, so that $\tau_{0}$ is the optical depth at 160~\um\ ($\tau_{160}$). We use a fixed spectral index $\beta = 2$, consistent with the grain models of \citet{Draine2003a}.
 
Since we have two observations and two unknowns, we are able to derive a unique solution for effective temperature $T$ and optical depth $\tau_{0}$.
The simplest solution can be found by making the optically thin ($\tau_{\nu} \ll 1$) approximation $(1 - e^{-\tau_{\nu}}) \approx \tau_{\nu}$.
Applying this approximation within our Equation~\ref{eq:dust_intensity}, measured intensity $\overline{I}_{i}$ in either band can be expressed

\begin{equation} \label{eq:dust_optthin}
   \begin{aligned}
       \overline{I}_{i} = BP_{i}\Big[ B_{\nu}(T)\, \tau_{\nu} \Big] = BP_{i}\Big[ B_{\nu}(T)\, \tau_{0} \big(\nu / \nu_{0} \big)^{\beta} \Big] \\
       = BP_{i}\Big[ B_{\nu}(T)\, \big(\nu / \nu_{0} \big)^{\beta} \Big]\, \tau_{0} 
   \end{aligned}
\end{equation}

Recalling that the bandpass function $BP_{i}$ is primarily an integral over frequency, the constant-in-frequency $\tau_{0}$ can be pulled outside of the function.
The ratio of the intensities in the two bands excludes the parameter $\tau_{0}$ entirely.

\begin{equation} \label{eq:dust_intensityratio}
    \frac{\overline{I}_{70}}{\overline{I}_{160}} = \frac{BP_{70}\Big[ B_{\nu}(T)\, \big(\nu / \nu_{0} \big)^{\beta} \Big]}{BP_{160}\Big[ B_{\nu}(T)\, \big(\nu / \nu_{0} \big)^{\beta} \Big]}
\end{equation}

This expression for the ratio of the measured intensities depends only on one parameter, the effective temperature $T$.
The expression is easily evaluated for a range of $T$, producing a series of modeled intensity ratio values.
Using this numerical grid, we interpolate from the observed intensity ratio values to temperatures.
With derived effective temperatures in hand, we rearrange our Equation~\ref{eq:dust_optthin} and evaluate the expression for $\tau_{0}$ using the measured intensities in one of the bands.

\begin{equation} \label{eq:dust_solvefortau}
    \tau_{0} = \frac{\overline{I}_{i}}{BP_{i}\Big[ B_{\nu}(T)\, \big(\nu / \nu_{0} \big)^{\beta} \Big]}
\end{equation}

We have numerically evaluated the above two expressions to derive the temperature and dust optical depth over the map. We have converted the calculated 160~$\mu$m optical depth into H-nuclei column densities (described below) and the results are shown in Fig.~\ref{fig:dust_mask}. In principle, there is no need to make the optically thin approximation.
We can work directly from Equation~\ref{eq:dust_intensity} and write the intensity ratio of the two bands without any approximation or cancelling of terms.
We can then use the calculated optical depth in the optically thin approximation to derive, from the rewritten intensity ratio, an improved temperature, and use that to numerically derive a new optical depth from Equation~\ref{eq:dust_intensity} and continue this iteration until convergence is achieved.
As the 160~$\mu$m optical depth in the shell is rather small, this procedure converges rapidly. Tests on a few points in the shell demonstrate that the iteration produces only small ($\sim$5\%) changes in the optical depth. As this is comparable to the calibration uncertainty, we have elected to continue the analysis with the optically thin approximation.

In order to convert the 160~\um\ optical depth to hydrogen nucleus column density, $N(H)$ = $N$(HI) + 2$N$(H$_2$), we use the \citet{Draine2003a} R$_{V} = 3.1$ value of the dust extinction cross section per hydrogen nucleus at 160~\um, $C_{\text{ext},160}/\text{H} = 1.9 \times 10^{-25}~\text{cm}^{2}/\text{H}$, and solve Equation~\ref{eq:dust_column} for $N(H)$.
The maps of dust temperature and $N(H)$ are presented in Figure~\ref{fig:dust_mask}.
\begin{equation} \label{eq:dust_column}
    \tau_{160} = (C_{\text{ext},160}/\text{H}) \times N(H)
\end{equation}

We assume a half-elliptical shell, with the major and minor axes lengths from Section~\ref{sec:shell_description}, and create a mask tracing the limb-brightened Eastern edge.
From the H nucleus column densities, which range from 0.3 to 1.3$\times 10^{23}$~cm$^{-2}$ along the shell, we calculate a total gas mass (excluding He) of this region, finding a value of 8.5$\times10^{3}$~\(\textup{M}_\odot\).

We consider this mass estimate an upper limit due to line-of-sight contribution.
This measurement picks up some other components of gas that are not part of the shell. One of the brightest lines-of-sight in $^{12}$CO (east of Wd2) is a superposition of the shell and ridge components, which can be disentangled in the CO emission but not in the dust. We find that the masses (estimated from $^{13}$CO using the same technique as explained in the later paragraphs of this section) along the brightest lines-of-sight in CO split roughly 60/40, shell/ridge. The mass corresponding to the ridge part comes out to be 297~\(\textup{M}_\odot\) putting an additional $<$ 5\% error on the shell mass.

Our gas mass estimate from the dust column also includes more diffuse foreground and background contribution, as the observed FIR intensity, and consequently the calculated column density, is non-zero even a few arcminutes away from the shell.
This is distinct from the non-shell components discussed above in that this diffuse contribution is larger scale, extending past the central RCW~49 region in Figure~\ref{fig:dust_mask}, and is probably physically distant from and thus completely unrelated to the shell.
We make a rough estimate of this combined ``background'' by sampling a column density of 1.3$\times 10^{22}$~cm$^{-2}$ a few arcminutes away from the shell.
If we subtract this background from the map and then carry through our previous calculation, we find that it may account for up to 30\% of the mass in our upper limit estimate.
Rather than make this rough background estimate and subtraction, we simply reiterate our interpretation of the mass measurement as an upper limit.

Finally, the half-ellipse model captures most of the visible \cii\ shell, but the shell extends slightly past the northern edge of the mask by about $20\degree$ in azimuth, and past the southern edge by about $10\degree$.
We make a rough correction for this by multiplying the mass estimate by $7/6$, assuming a constant linear density along the edge of the shell.
It is possible to make a more articulated shell mask for a more precise measurement, but this would necessitate more complex assumptions. \\

The second method to estimate the gas mass is by determining the C$^+$ column density $N$(C$^+$). As detailed in Appendix~\ref{sec:cii_opacity_appen}, the \13cii\ line shows that optical depth effects are small over most of the \cii\ arc, except for the brightest emission spot. Hence, we have calculated the C$^+$ column density in the optically thin limit following \citet[equation~A.1]{2018A&A...615A.158T}.
We assume an excitation temperature $\sim$ 100~K (lower limit), which is a reasonable kinetic gas temperature in the \cii\ emitting layer of a PDR to excite the C$^+$ from $^2P_{1/2}$ $\to$ $^2P_{3/2}$. We calculated the column density (upper limit) of the observed limb-brightened part for a region similar to the mask used in the dust mass estimation and for the emission within -25 to 0~km~s$^{-1}$. We found an average $N$(C$^+$) $\sim$ 7.2 $\times$ 10$^{18}$~cm$^{-2}$. The column density $N$(C$^+$) is not very sensitive to the choice of $T_{ex}$. For instance, assuming $T_{ex} = 200$~K instead of 100~K decreases the calculated $N$(C$^+$) by 19\%.
  
A pixel-by-pixel sum of $N$(C$^+$) over the entire thickness allowed us to estimate the H gas mass. Using the abundance ratio of C/H = 1.6 $\times$ 10$^{-4}$ \citep{2004ApJ...605..272S}, we get the corresponding
H gas mass\footnote{We note that our analysis assumes a purely atomic hydrogen column density but the C$^+$ arises in both the warm atomic and molecular gas.  Since the collisional excitation rate of C$^+$ by H2 is $\sim$ 0.7 times the rate of excitation by
atomic hydrogen \citep{2014ApJ...780..183W} we expect that a purely H$_2$ column would have $\sim$ 1.5 times the mass.} $\sim$ 4.6 $\times$ 10$^3$~\(\textup{M}_\odot\), which is very similar to the mass calculated from the dust emission. \\

Finally, we can estimate the H$_{2}$ gas mass from $N$($^{13}$CO), which requires determination of its excitation temperature, $T_{\rm ex}$. Again, these estimations were done for the same masked region as described above and for the emission within -25 to 0~km~s$^{-1}$. As detailed in Appendix~\ref{sec:co_tex_colden_appen}, we get an average $T_{\rm ex}$ $\sim$ 14.5 $\pm$ 1~K and an average $N$($^{13}$CO) $\sim$ (8.8 $\pm$ 0.1) $\times$ 10$^{15}$~cm$^{-2}$ such that the average $N$($^{12}$CO) $\sim$ 4.6 $\times$ 10$^{17}$~cm$^{-2}$, using $^{12}$CO/$^{13}$CO = 52 \citep{2005ApJ...634.1126M}. Similar to the calculation of H gas mass, by summing pixel-by-pixel the $N$($^{13}$CO) and by using $^{12}$CO/H$_{2}$ = 8.5 $\times$ 10$^{-5}$ \citep{2010pcim.book.....T}, we get an H$_2$ gas mass of $\sim$ 1.5 $\times$ 10$^{3}$~\(\textup{M}_\odot\). As expected from the more fragmented CO emission, this mass estimate is somewhat less than derived from the dust or the \cii\ emission. \\ 

In summary, our mass estimate using the dust includes both the atomic and molecular gas. The estimate using \cii\ emission (CO dark gas) depends on the
molecular fraction but differs by a factor of only 1.5 between pure atomic and pure molecular columns. The estimate from $^{13}$CO is for the molecular gas alone. Thus the shell mass estimated from dust (8.5 $\times$ 10$^{3}$~\(\textup{M}_\odot\)) is comparable to that estimated by \cii\ and $^{13}$CO together (6.1 $\times$ 10$^{3}$~\(\textup{M}_\odot\)). The range of masses obtained through the different methods is about a factor of 1.4 in mass. \\

In order to estimate the entire shell's mass, we assume that the shell occupies the space between two concentric prolate spheroidal shells with North-South semimajor axes $a=$ 6.5 and 7.5~pc and semiminor axes $b=$ 3.5 and 4.5~pc such that the shell thickness is about 1~pc.
We also assume that the limb-brightened region is represented by the volume of a prolate spheroid of $a=$ 7.5~pc and $b=$ 4.5~pc from which the volume of an elliptic cylinder of $a=$ 6.5~pc and $b=$ 3.5~pc is subtracted, which we call a ``cored spheroid''.
The conversion from the volume of a cored spheroid to that of a spheroidal shell gives us a geometric correction factor of 2.5, which we determined numerically.
By a simple argument of symmetry, this correction is valid for our quarter-spheroid shell assumption.
Applying that factor to our limb-brightened part's mass (derived from dust), and including the corrective factor of $7/6$ explained earlier, we get the corrected shell mass estimate $\sim2.5\times10^{4}$~\(\textup{M}_\odot\). \\

\begin{table*}

\centering
\begin{threeparttable}
\caption{Densities, temperatures and pressures calculated for the shell of RCW~49.}
\begin{tabular}{l c c c c c}
\hline\hline
\noalign{\smallskip}
 Region & $n$ (cm$^{-3}$) & $T$ (K) & $p_{\rm th}/k$ (cm$^{-3}$~K) & $p_{\rm rad}/k$ (cm$^{-3}$~K) & $p_{\rm turb}/k$ (cm$^{-3}$~K) \\
  \hline
 \noalign{\smallskip}
Plasma\tnote{a} & 0.71 & 3.13 $\times$ 10$^6$ & 4.9 $\times$ 10$^{6}$ & - & -  \\
Ionized gas\tnote{b} & 317 & 7.7 $\times$ 10$^3$ & 4.9 $\times$ 10$^6$ & - & -\\ 
PDR/\cii\ layer\tnote{c} & 4 $\times$ 10$^{3}$ & 300 & 1.2 $\times$ 10$^6$ & 2.6 $\times$ 10$^6$ & 5.9 $\times$ 10$^6$ \\
 
 \hline
\end{tabular}

Notes: Columns from left to right are region, H density, temperature, and thermal, radiation, and turbulent pressures. 
\begin{tablenotes}
\item[a]\small $n$ is abundance of e$^-$ from ionization of H and $p_{\rm th}/k$ = $2.2nT$, where the factor 2.2 accounts for ionization of H and singly
ionized He.
\item[b]\small $n$ is e$^-$ density from ionization of H and $p_{\rm th}/k$ = $2nT$, where factor 2 accounts for ionized H$^+$, and neutral He.
\item[c]\small $n$ is H density and $p_{\rm th}/k$ = $nT$. The radiation pressure $p_{\rm rad} = L_{\rm bol}/4{\pi}kR^2c$, where  $R$ is the radius of the shell. The turbulent pressure $p_{\rm turb} = \mu mn ~\Delta v_{\rm turb}^2/8\ln{2}~k$, where $\mu =1.3$ is the mean molecular weight, $m$ is hydrogen mass and $\Delta v_{\rm turb}^2 = \Delta v_{\rm FWHM}^2 - (8\ln{2}~kT/m_{\rm c})$, where $m_{\rm c}$ is the carbon mass.     
\end{tablenotes}

 \label{pressures}
 \end{threeparttable}
\end{table*}

We can check our assumption of the shell mass against observed extinction through the foreground gas and shell.
If we assume the gas distribution is uniform throughout the shell, then we can divide the total shell mass by the surface area of a quarter of a spheroidal shell at its mean semimajor and semiminor axes $a=$ 7~pc and $b=$ 4~pc and then convert the surface mass density to $A_{V}$ using the factor $N(H)/A_{V} = 1.9\times 10^{21}~\text{cm}^{-2}$ \citep{Bohlin_NH_to_Av} for an $R_{V} = 3.1$ reddening law.
Including the geometric factor of 2.5 (but excluding the $7/6$ factor so that we match our surface area assumption), the mass estimate from far-infrared dust emission results in an extinction of $A_{V,\,shell} \sim 18$ through the shell.

\citet{VA2013} and \citet{VPHAS_Wd2_2015} both measured an average $A_{V} \sim 6.5$ towards Wd2 cluster members, each finding the reddening to be $R_{V} \sim 3.8$.
\citet{Hur2015} reported an abnormally high reddening law, $R_{V} = 4.14$, towards early-type cluster members and a more typical law, $R_{V} = 3.33$, towards foreground stars in the same field.
They report a total $E(B-V) \approx$ 1.7 to 1.75 for most cluster members, and find a foreground $E(B-V)_{fg} \approx 1.05$, which suggests $A_{V,\,fg} \sim 3.5$.
\citet{2015AJ....150...78Z} report $E(B-V) \approx$~1.8 to 1.9 based on reddening of line emission from ionized gas.

All four of these measurements are consistent with $A_{V} \sim 6.5$ towards Wd2 cluster members, implying $A_{V,\,shell} \sim 3$ towards the cluster after accounting for the foreground extinction measured by \citet{Hur2015}.
This suggests a significantly thinner shell than the $A_{V,\,shell} \sim 18$ we predict from our shell mass estimate, assuming a uniform shell.
We explain this discrepancy by proposing that the thickness of the shell varies significantly across its surface, a claim further supported by the fragmented CO distribution we observe across the shell and the higher $A_{V} \sim 15$ fit by \citet{2004ApJS..154..315W} to two young stellar objects embedded in the western shell.
If we picture an optical path originating from the cluster and extending east, passing through the bright eastern shell, this path experiences $A_{V} \sim 18$ extinction according to our mass calculations.
But according to the optical extinction measurements, the thickness of the shell drops dramatically as the optical path sweeps towards us.
In fact, we do not detect any bright shell component along the line of sight towards the cluster in the \cii\ and CO spectra and pv diagrams.
We draw the conclusion that the extinction associated with the cluster probes a much thinner section of the shell than the limb-brightened eastern shell from which we extrapolate our mass estimate, and so our geometrically-extrapolated mass measurement must be considered an upper limit.

One could imagine combining our limb-brightened shell mass estimate with an optical extinction map like those shown in Figure~14 in the paper by \citet{Hur2015} or Figures 8, 9, and 25 in the paper by \citet{2015AJ....150...78Z} in order to understand the variation in thickness across the shell.
High confidence in the 3-dimensional positions of the cluster members and the geometry of the shell would be required in order for such a measurement to have any meaning, and this is beyond the scope of the present paper.

\subsubsection{Energetics} \label{sec:energetics}
Using the total mass (excluding He) of the shell $\sim$ 2.5 $\times$ 10$^4$~\(\textup{M}_\odot\) and its expansion velocity $\sim$ 13~km~s$^{-1}$, we calculated its kinetic energy, $E_{\rm kin}$ $\sim$ 4 $\times$ 10$^{49}$~ergs. \\

To assess the contribution from stellar winds in driving the shell, we need to estimate the stellar wind energy of Wd2. 
The early-type cluster members are catalogued along with their established or estimated stellar types by \cite{TFT2007}, \cite{VA2013}, and \cite{VPHAS_Wd2_2015}. We used the theoretical calibrations of \cite{Martins2005} to estimate effective temperature $T_{\text{eff}}$, surface gravity log$~g$, and luminosity $L$ from the spectral type. For WR20b (WN6ha; \citealt{vanderHucht2001}) and each component of the WR20a binary (WN6ha+WN6ha), we assume parameters fitted to WR20a by \cite{Rauw2005}. See Appendix~\ref{sec:wd2_appendix} for additional details about the synthesized catalog, the measurements derived from the catalog, and the uncertainties on those measurements. 

\begin{figure*}[htp]
\centering
\includegraphics[width=0.95\textwidth]{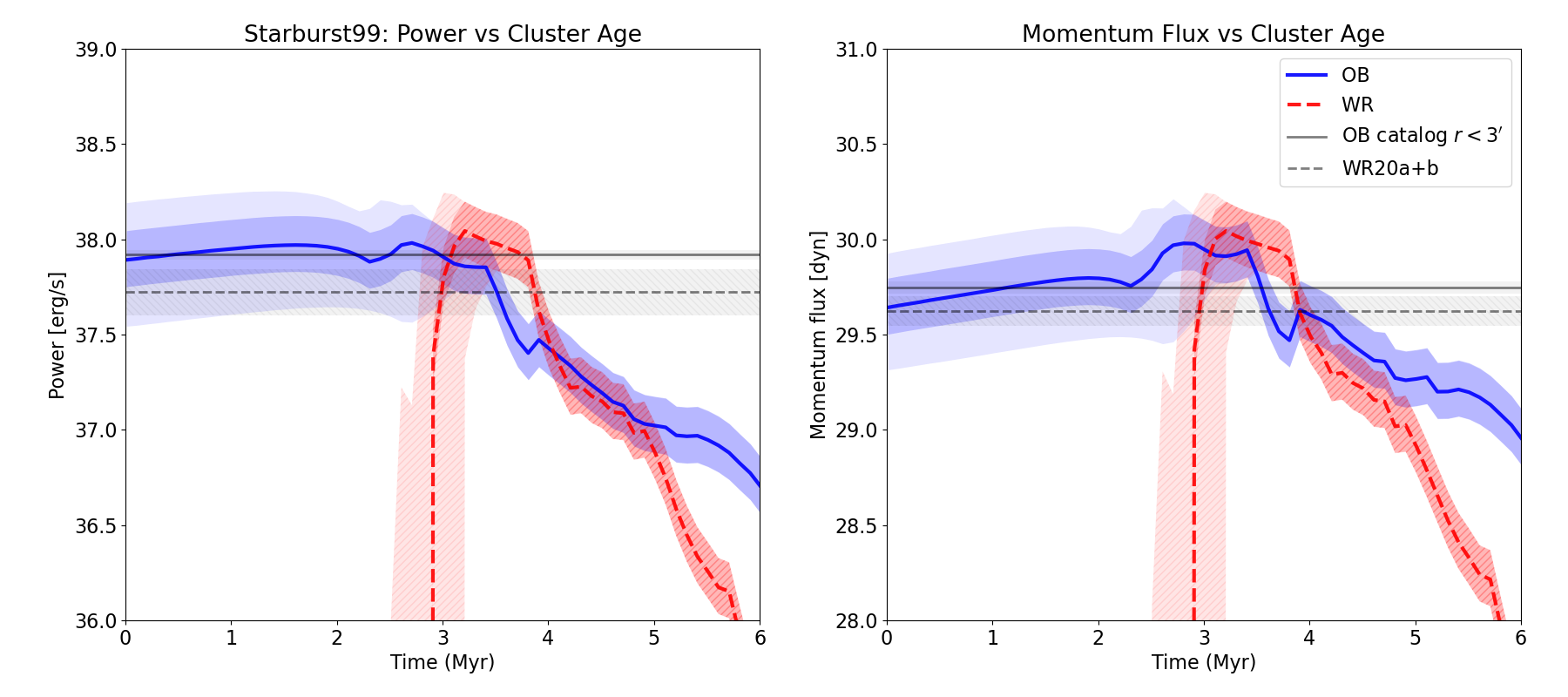}
\caption{\label{fig:sb99} \texttt{Starburst99} predictions of the mechanical luminosity (left panel) and momentum transfer rate (right panel) due to the stellar winds. The \texttt{Starburst99} simulation configurations are described in Appendix~\ref{sec:sb99_appendix}. The solid blue and dashed red lines, and associated solid shaded blue and hatched red regions, give the values for OB and WR stars, respectively. The horizontal lines, solid and dashed and their associated shaded regions, mark the values calculated from the observed OB and WR stars as described in Section~\ref{sec:energetics} as well as in Appendix~\ref{sec:wd2_appendix}. The darker-shaded regions reflect uncertainty in the total cluster mass, as described in the Appendix~\ref{sec:sb99_appendix}. The lighter-shaded regions reflect total cluster mass uncertainty as well as uncertainty in the maximum stellar mass; the upper limit uses a [1, 120]~$M_{\odot}$ range, and the lower limit uses a [1, 80]~$M_{\odot}$. Note the effect of the maximum stellar mass on the age at which WR stars appear in these models.}
\end{figure*}

The combined mass loss rate, mechanical energy injection ($1/2~\dot M v_{\infty}^{2}$), and momentum transfer rate ($\dot M v_{\infty}$) by the O and B stars within $3'$ of the cluster center, at the peak of the X-ray emission \citep{2019ApJS..244...28T}, is $(3.1 \pm 0.2) \times 10^{-5}$~\(\textup{M}_\odot\)~yr$^{-1}$, $(8.3 \pm 0.5) \times 10^{37}$~ergs~s$^{-1}$, and $(5.6 \pm 0.4) \times 10^{29}$~dyn \citep{Leitherer2010}.
The WR binary WR20a contributes an additional $1.9^{+0.25}_{-0.20}\times 10^{-5}$~\(\textup{M}_\odot\)~yr$^{-1}$, $3.6^{+1.3}_{-1.1} \times 10^{37}$~ergs~s$^{-1}$, and $3.0^{+0.70}_{-0.60} \times 10^{29}$~dyn.
Using the evolutionary spectral synthesis software \texttt{Starburst99} \citep[see description in Appendix~\ref{sec:sb99_appendix}]{Leitherer2014Starburst99}, we estimate that over the lifetime of the cluster ($\sim$2~Myr), the OB stars have injected $\sim 6 \times 10^{51}$~ergs via their winds (see Fig.~\ref{fig:sb99} and Appendix~\ref{sec:sb99_appendix} for additional detail).

WR stars represent a rather short-lived phase of the lifetimes of very massive stars, forming after about $\sim$2 to 3~Myr, depending on their mass, and lasting a few hundred thousand years if we use the results of \texttt{Starburst99} as a guide.
The components of WR20a are estimated by \citet{Rauw2005} to be $\sim80~\text{M}_{\odot}$ each, the most massive observed stars in the cluster; if this is close to their initial masses, then neglecting binary effects on their evolution, \texttt{Starburst99} would suggest that they formed after $\sim3$~Myr, creating some tension with most of the age estimates of the cluster.
If these Wolf-Rayet stars originated as much more massive objects, like the $\sim126~\text{M}_{\odot}$ predicted by \citet{Ascenso2007} to be the most massive star in the cluster based on their assumed IMF, then \texttt{Starburst99} would suggest that WR stars could have formed after $\sim2$~Myr, which agrees better with independent age estimates.

\citet{Rauw2005} observe evidence of enhanced surface hydrogen abundance in the components of WR20a, indicating that they (and WR20b, if we assume the components are identical) are still in the core hydrogen-burning phase and, based on their position in the Hertzsprung-Russell diagram, are only $\sim1.5$~Myr old and have present-day masses very similar to their initial masses.
In any case, over the lifetime of the cluster, the OB stars will have dominated the kinetic feedback; but over the last 2 to 3$\times 10^{5}$ years, the winds of the WR binary alone will have contributed $\sim4 \times 10^{49}$~ergs.

\begin{figure*}[htp]
\centering
\includegraphics[width=130mm]{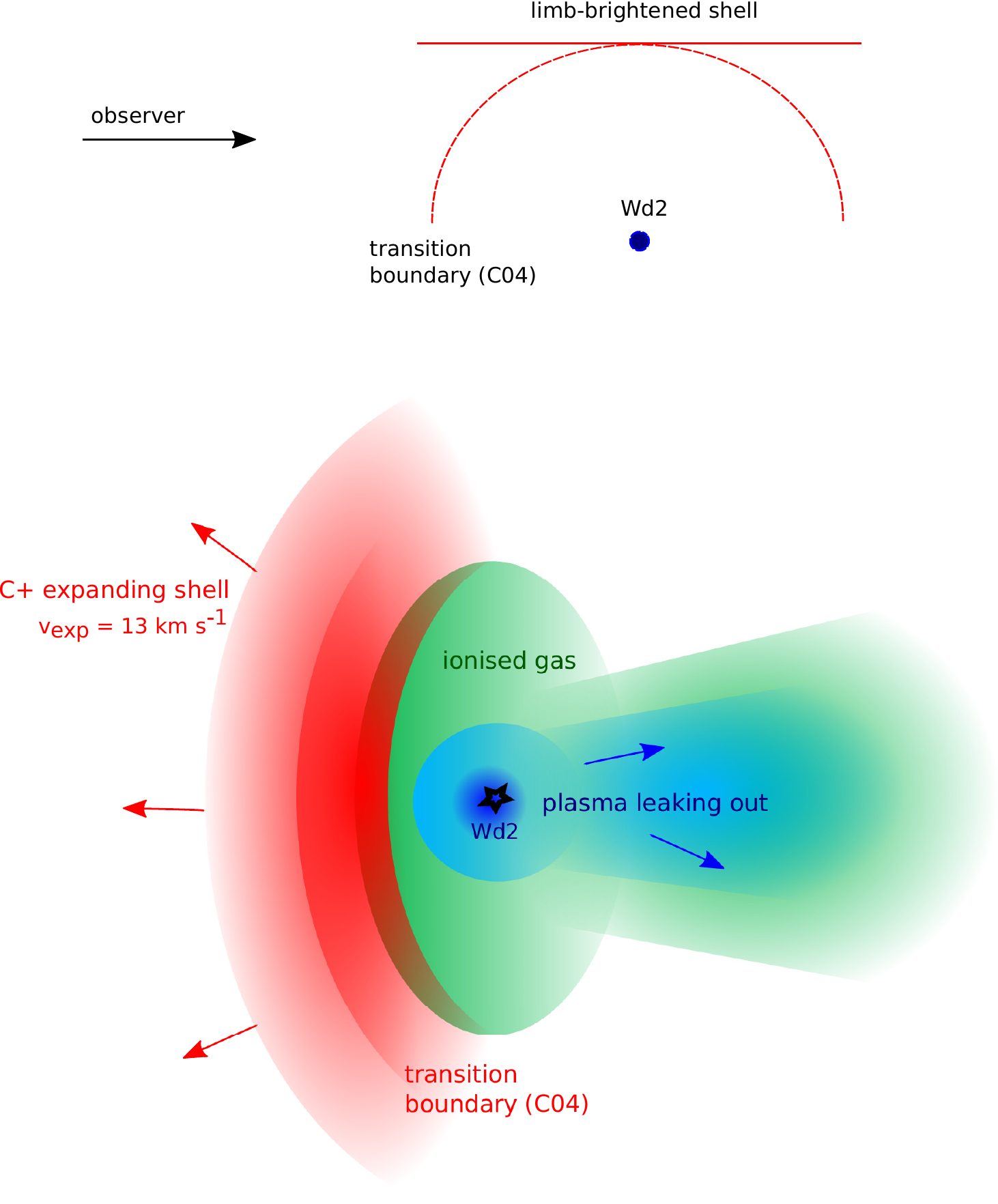}

\caption{2D (upper panel) and 3D (lower panel) representations of RCW~49's shell as seen by the observer. The transition boundary from \citet[Fig.~1]{2004ApJS..154..322C} is marked and labelled with `C04'.
The plasma (in blue), ionized gas (in green) and the PDR (in red) are shown. The limb-brightened part of the shell in the 2d illustration is actually the expanding \cii\ shell seen in the 3D diagram. The transition boundary overhangs from the \cii\ shell into the ionized gas structure behind it.}

\label{geometry} 
\end{figure*}

Given the similarity in spectral characteristics, the contribution by WR20b is likely similar to that of a single component of WR20a (approximately half the values given above).
However, WR20b is significantly offset ($>3' \sim 3.5$~pc) from the center of the X-ray emission tracing the $\sim3$~pc radius plasma bubble, so it is probably not playing as direct a role as WR20a in powering this bubble. \\

The importance of the plasma's thermal energy in expanding the shell can be assessed by comparing thermal pressures, $p_{\rm th}$, in the hot plasma, ionized gas and PDR of RCW~49.
To characterize the hot plasma, we use the RCW~49 \textit{Chandra}/ACIS observations reported by \citet{2019ApJS..244...28T}.
These authors fit X-ray spectra towards RCW~49 with plasma models using the spectral fitting software Xspec \citep{1996ASPC..101...17A}.
The plane-parallel, constant temperature shocked plasma model labeled \textit{pshock2} represents the diffuse plasma towards the center of Wd2, so we take from this model the fitted temperature and surface emission measure listed in the Table~5 by \citet{2019ApJS..244...28T}.
They independently fit spectra from the inner region towards Wd2 and the outer region out to $\sim$3 arcminutes away (see their text for details), and label these independent fits ``Wd2~inner'' and ``Wd2~outer''.
We take the \textit{pshock2} parameters of the ``Wd2~outer'' region and assume the plasma fills a sphere whose circular cross-section equals the observed areas of ``Wd2~outer'' and ``Wd2~inner'' combined, yielding a radius of 2.7~pc.

Following the calculations of \citet{Townsley2003}, we obtain the electron density (assuming a line-of-sight distance $2r$ through the plasma) and temperature, which are listed in Table~\ref{pressures}.
Finally, assuming the temperature and pressure are constant throughout the spherical bubble, the total thermal energy of the plasma is about $2.4 \times 10^{48}$~ergs. There is clearly a mismatch (lower) in the thermal energy of the hot plasma, the mechanical energy injected over the lifetime of the stellar cluster, and the kinetic energy of the expanding shell. We will revisit this in Sect.~\ref{sec:morph}. 

We followed the work by \cite{2015ApJ...813...24P} to estimate the temperature and density in the ionized gas in the \hii\ region using their observed H109$\alpha$ line properties. We excluded from the calculation their ``region~B'' due to its outlying line-of-sight velocity, which indicates it may be from the blue-shifted foreground ridge component we detect in \cii. We assumed a hollow spherical geometry with an outer radius of 5~pc, bordering the PDR, and an inner radius of 2.7~pc, bordering the X-ray emitting plasma, and calculated the electron density and temperature (listed in Table~\ref{pressures}).

Lastly, for the PDR, we compared our observations with existing PDR models (PDR Toolbox; \citealt{2006ApJ...644..283K}, \citealt{2008ASPC..394..654P}). To use these, we needed an estimation of the FUV flux incident on the PDR. We used the $T_{\rm eff}$ and log $g$ (or appropriate WR parameters from \citealt{Rauw2005}) for each catalogued star to select models from the PoWR stellar atmosphere grids \citep{PoWRCode3}.
From the synthetic spectra provided by the PoWR models, we integrated the total flux between 6 to 13.6~eV. Using the stellar coordinates and these FUV fluxes from all stars within 12\arcmin\ (including WR20a and WR20b, though these contribute only a few percent of the total FUV flux), we estimated the integrated FUV flux between 6 to 13.6~eV, often expressed as $G_{0}$ in terms of the Habing field, to be $\sim$ 2--3 $\times~10^{3}$ in Habing units at the limb-brightened shell radius of $\sim$ 5--6 pc from Wd2. This value should be considered an upper limit, as the extinction between the illuminating cluster and the PDR due to dust in the \hii\ region has not been accounted for. For the average intensities (within the masked region shown in Fig.~\ref{fig:dust_mask}) of \cii\ ($\sim$ 112~K~km~s$^{-1}$) and $^{12}$CO ($\sim$ 41~K~km~s$^{-1}$) at a FUV radiation field, $G_{\rm 0}$ $\sim$ 10$^{3}$ in Habing units, we determined the PDR's H density and temperature using the models from the PDR toolbox\footnote{http://dustem.astro.umd.edu/} \citep{2006ApJ...644..283K,2008ASPC..394..654P}. Using our observed line ratio of \cii\,/$^{12}$CO(3-2) (after conversion from K~km~s$^{-1}$ to erg~cm$^{-2}$~s$^{-1}$~sr$^{-1}$) and comparing with the modeled line ratio\footnote{http://dustem.astro.umd.edu/models/wk2006/ciico32web.html} as a function of the cloud density and $G_{\rm 0}$, allowed us to estimate the density. Further, using this density and the $G_{\rm 0}$, we constrained the PDR temperature using the modeled PDR surface temperature map\footnote{http://dustem.astro.umd.edu/models/wk2006/tsweb.html}. The temperature is not strongly dependent on $G_{\rm 0}$ in the derived density range.

A list of the derived densities, temperatures and thermal pressures is presented in Table~\ref{pressures}.

Using the sum of the bolometric luminosities, $L_{\rm Bol}$, for the OB stars and the WR20a star, we can estimate the radiation pressure (as mentioned in Table~\ref{pressures}). The bolometric luminosities of the OB stars within 3$\arcmin$ were taken from the spectral type calibrations of \citet{Martins2005} as described above and \citet{Rauw2005} provides the bolometric luminosities of the components of WR20a based on their fit to its spectrum.

We can also estimate the turbulent pressure, $p_{\rm turb}$ (in Table~\ref{pressures}), from the full width half maximum of the observed \cii\ emission line profile. These results reveal that there is rough pressure equipartition between the thermal, turbulent and radiation pressure in the PDR. Examining Table~\ref{pressures}, we conclude that the hot plasma, the ionized gas and the PDR are in approximate pressure equilibrium as expected for a stellar wind shell driven by mechanical energy input from the central star cluster \citep{1977ApJ...218..377W}.

\section{Discussion}

\subsection{Morphology of the shell and the role of WR20a in its expansion} \label{sec:morph}
As discussed in Sect.~3.3.2, the kinetic energy of the expanding shell is much higher than the thermal energy of the plasma. We emphasize that the mechanical luminosity of the stellar cluster well exceeds the requirements for driving the shell. Therefore, we surmise that much of the thermal energy was lost once the shell broke open to the west and the hot gas expanded freely into the environment (as shown in Fig.~\ref{geometry}). Beside the adiabatic cooling associated with this ``free'' expansion, evaporation of entrained cold gas into the hot plasma due to electron conduction (\citealt{1977ApJ...218..377W}; \citealt{1977ApJ...211..135C}) may have led to regions that are dense and cool enough to allow rapid cooling, resulting in a rapid loss of thermal energy. 

We also recognize that there is a timescale issue. The shell radius of 6~pc and the expansion velocity of 13~km~s$^{-1}$ imply an expansion timescale of $\sim$ 0.5~Myr. For an enclosed bubble driven by adiabatic expansion of the hot plasma created by a continuous input of mechanical energy by stellar winds, the expansion time scale is only 0.27~Myr (using equations~51 and 52 of \citealt{1977ApJ...218..377W}).
In contrast, the age of the Wd2 cluster is $\sim$2~Myr according to most age estimates. It should be noted that a different choice of the heliocentric distance (say up to 8~kpc) of RCW~49 would increase the estimated expansion timescale of the shell to 0.52~Myr, which is still inconsistent with the age of Wd2. Hence, the average expansion velocity over most of the cluster lifetime must have been $\lessapprox$ 2~km~s$^{-1}$ and only very recently ($\lessapprox$ 0.2~Myr), the shell has been accelerated to 13~km~s$^{-1}$.
We infer that feedback from OB stars initially drove shell formation and expansion but that this bubble quickly burst, releasing the hot plasma. At that point, expansion would rapidly slow down due to continued sweeping up of the cold gas in the environment.
The recent re-acceleration of the shell might be connected to the evolution of the most massive stars (WR20a and 20b) to the Wolf-Rayet phase. If we assume the bubble has burst, expansion must be driven by momentum transfer.
The issue is that the Wolf-Rayet stars do not seem to inject significantly more momentum than the ensemble of OB stars; this issue arises in the momentum transfer rates calculated directly from the observed WR stars and their properties as well as those more generally predicted through \texttt{Starburst99} simulations. We do not propose any solutions to this conundrum, at present; this requires further detailed analyses of the member stars, especially WR20a and 20b, as well as the expanding shell.

\subsection{Previous studies and larger scale structure}

Our picture of a single shell at the center of RCW~49 differs from that of the two shell (separated by the ridge) scenario presented by \citet{1997A&A...317..563W} and \citet{2013A&A...559A..31B}.
Owing to the kinematic information provided by the high spectral resolution of the \cii\, data that the radio data lacks, we were successful in decoupling the central ``ridge'' from the shell.
The ``radio ring B'' surrounding WR20b appears to be a superimposition of filamentary structures, rather than a coherent ring. These structures generally extend toward the main Wd2 cluster, suggesting that Wd2, rather than WR20b, dominantly influences their morphology.
We find that the ridge, too, is a superimposition of hot gas and dust components well-separated in velocity space, which may explain the variation in H137$\beta$ and H109$\alpha$ radio recombination line velocities observed by \citet{2013A&A...559A..31B} and \citet{2015ApJ...813...24P}, respectively.
Some red-shifted sections of the ridge seem to be connected in velocity space to a larger $+16$ km/s molecular cloud \citep{2009ApJ...696L.115F}, which may indicate that these sections lie beyond the cluster and limb-brightened shell.

While we argue that the expansion of the shell is driven by stellar winds of WR20a, on a larger scale the
observed velocity structure (blue and red-shifted components of gas) in RCW 49 could be guided by the dynamics of several individual molecular clouds which predate Wd2. \citet{2009ApJ...696L.115F} suggests that a collision between two of these clouds may have contributed to the
formation of Wd2 and, consequently, RCW~49. Additionally, \citet{2019ApJS..244...28T} observed diffuse hard X-ray emission
from far-west of Wd2 toward a pulsar wind nebula that is indicative of a cavity supernova remnant, suggesting an earlier generation of massive star formation in RCW~49. Perhaps this earlier generation of star formation is responsible for the large scale velocity dispersion observed by \citet{2009ApJ...696L.115F}, while the local shell expansion that we present in this work is driven by the stellar winds of WR20a.

\citet{2004ApJS..154..315W} studied star formation in different regions of RCW~49 using the GLIMPSE survey and found that most of the star formation is occurring within a 5 pc radius (similar to the transition boundary of \citealt{2004ApJS..154..322C}) from the Wd2 cluster. At larger distances, a second generation of star formation perhaps triggered by Wd2 is also suggested, based on the massive (B2--3) young stellar objects (YSOs) detected. To put it in context of our findings, this implies that star formation is occurring mainly in the ridge of RCW 49, while a second (younger) generation of star
formation is probably triggered in the shell. While the former one may reflect the cloud-cloud collision event highlighted by \citet{2009ApJ...696L.115F}, the latter is likely triggered by
feedback from the Wd2 cluster. However, \citet{Hur2015} suggests triggered star formation from the radiative feedback of Wd2 as a possible explanation for the enhanced abundance of pre-main sequence (PMS) candidates observed in the ridge in X-ray by \citet{Naze2008_Xray}. We suggest both the cloud-cloud collision event and the compression from Wd2's radiative feedback could be a cause for the triggered star formation in the ridge.

\citet{2004ApJS..154..315W} reported a total of $\sim$ 7000 YSOs in RCW~49 with a total mass of 4500~\(\textup{M}_\odot\) and we infer that a fraction of it constitutes triggered star formation in the shell. Follow-up studies in X-ray and IR wavelengths can shed more light on the triggered star formation efficiency of the shell.

The stellar mass of the Wd2 cluster is $\sim$ 3.1 $\times$ 10$^4$~\(\textup{M}_\odot\) for stars with masses $<$ 0.65~\(\textup{M}_\odot\) and $\sim$ 4 $\times$ 10$^4$~\(\textup{M}_\odot\) for stars with masses $>$ 0.65~\(\textup{M}_\odot\) \citep{Zeidler2017}. This means that the stellar mass due to triggered star formation in the shell is smaller than the mass of Wd2 cluster. Furthermore, from their modeling results that calculate emission from envelopes, disks, and outflows surrounding stars, \citet{2004ApJS..154..315W} found the most massive YSO in RCW~49 is $<$ 5.9~\(\textup{M}_\odot\), suggesting that the new generation of stars will be relatively lower in mass compared to the stars in Wd2 and feedback from this next generation of stars is expected to be limited. 



\subsection{Comparison with the shell of Orion}

The Orion Molecular Cloud (OMC) is the closest massive star-forming region that has been studied extensively in a wide range of wavelengths. OMC~1 is its most massive core associated with the well known HII region of M42 (the Orion Nebula). \citet{2019Natur.565..618P} reported an expanding shell driven by the stellar winds of the O7V star $\theta^1$ Ori C in the Orion Nebula. The velocity of the expanding Orion veil shell is similar to that of the shell of RCW~49, i.e. 13~km~s$^{-1}$, but has a mass of $\sim$ 2600~\(\textup{M}_\odot\), which is about 9 times lower than the mass (and also the kinetic energy) of the shell of RCW~49. The difference between the kinetic energies reflects the fact that the Orion veil shell is the result of the mechanical energy input from one O7V star while the shell in RCW~49 is the effect of a rich stellar cluster. Furthermore, Orion with an age of 0.2~Myrs is relatively younger compared to RCW~49, that has an age of at least 2~Myrs. Despite being created by a larger mechanical input, the shell of RCW~49 is moving at a similar velocity as that of the Orion veil. Perhaps this is because the shell of RCW~49 is broken toward the west and is venting out plasma, while the Orion veil seems to be a complete shell. However, likely, the veil will burst soon as well, releasing the hot plasma and hence the driving force of the expansion.

The O7V star $\theta^1$ Ori C lies at the front-side of OMC 1 and the shell expansion toward the rear is stopped by the dense core. In contrast, in RCW 49, backside of the bubble seems to have broken and the large scale molecular cloud in the velocity range of 11 to 21~km~s$^{-1}$ (as reported by \citealt{2009ApJ...696L.115F}) partially blocks the expansion of the red-shifted gas. Another interesting difference between the two shells is that we observe CO emission toward the same line-of-sight of the RCW~49's shell and its spatial distribution, though fragmented, outlines the shell (as in Figs.~\ref{co-chan} and \ref{13co-chan}), while the Orion veil shell lacks CO emission \citep{2020A&A...639A...2P}. Non-detection of CO in the Orion veil shell is attributed to the rather limited column density, $A_{\rm v}$ $\sim$ 2~mag, which corresponds to a gas column of $N$(H) = 4 $\times$ 10$^{21}$~cm$^{-2}$ \citep{2020A&A...639A...2P}, while we derived a maximum $A_{\rm v}$ $\sim$ 18 for RCW~49, which corresponds to a gas column of $N$(H) = 3 $\times$ 10$^{22}$~cm$^{-2}$. Existence of larger scale (and perhaps older) molecular clouds in RCW~49 has been established \citep{2009ApJ...696L.115F}. The old age of RCW~49 (2~Myrs) as compared to Orion (0.2~Myrs) puts RCW 49 at an advanced stage of evolution and more pronounced effects of stellar feedback in shaping its environment by sweeping dense molecular clouds seen as clumps toward the shell. Moreover, despite having a larger mechanical input, the denser gas column environment of RCW~49 could be another reason for its shell's expansion velocity to be similar to that of the Orion veil. The larger statistical study of the effects of stellar feedback in Galactic star forming regions initiated by the SOFIA FEEDBACK legacy program can help illuminate whether swept shells typically resemble Orion or RCW~49.

\subsection{Our understanding of stellar feedback}

This study of the expanding shell and molecular clouds in RCW~49 contributes toward our understanding of the stellar feedback in our Galaxy. In Sect.~3.6.2 we find that the evolution of the hot plasma, the \hii\ region and the PDR seems to be dominated by the energy injection by stellar winds of massive stars. 

Following our discussion on the next generation of star formation in Sect.~4.2, the total mass available for (triggered) star formation in the swept up shell is $\sim$ 10$^4$~\(\textup{M}_\odot\), which can be compared to the molecular clouds (of total mass $\sim$ 2 $\times$ 10$^5$~\(\textup{M}_\odot\)) from which Wd2 was formed \citep{2009ApJ...696L.115F}. As, even in a dense core, the star formation efficiency is less than unity, the total mass of the triggered cluster will be considerably less than that of Wd2. Moreover, as the mass of the most massive star in a cluster scales with the mass of the cluster \citep{1997ApJ...476..144M} and feedback can be expected to scale with the mechanical luminosity injected by the (most massive) star. Therefore, because each successive triggered star cluster will have lower mass than the previous, resulting in lower feedback, the triggered star formation process will gradually decrease.


Furthermore, comparing RCW~49 with Orion, we surmise that the effects of the stellar winds are limited to the earliest phases of the expansion and that, once the swept up shell breaks open, the hot gas is vented into the surroundings, the expansion stalls but if the stars are massive enough to enter the Wolf-Rayet phase, the expansion can be rejuvenated.

\section{Conclusions}

We presented for the first time large scale velocity integrated intensity maps of $^2$P$_{3/2}$ $\to$ $^2$P$_{1/2}$ transition of \cii\,, $J$ = 3 $\to$ 2 transition of $^{12}$CO and $^{13}$CO toward RCW~49. By analyzing the observed data in different velocity ranges, we successfully decoupled an expanding shell associated with RCW~49 from the entire gas complex. With the accessibility of better resolution data compared to the earlier studies (as discussed in Sect.~4.2) done toward RCW~49, we justified the presence of a single shell instead of previously thought two and characterised it for the first time. We find that the shell expanding toward us at $\sim$ 13~km~s$^{-1}$ is $\sim$ 1~pc thick and has a radius of $\sim$ 6~pc. We used dust SEDs and the column densities of \cii\, and $^{13}$CO, to estimate the mass of the shell $\sim$ 2.5 $\times$ 10$^4$~\(\textup{M}_\odot\). We quantified and discussed the effects of the stellar wind feedback, which mechanically powers the expansion of the shell of RCW~49. We constrained the physical conditions of the hot plasma and the ionised gas using the previous X-ray and radio wavelength studies, while using our new \cii\ and CO observations to determine the PDR parameters. Building on the geometry of RCW~49 derived from the dust emission studies, we put forward a 3D representation of the shell as seen by the observer, where the \cii\ shell overhangs the transition boundary between the ionised gas and the PDR. Based on the energy and time scale estimations, we suggest that the shell, initially powered by Wd2, broke open in the west releasing the hot plasma and its observed re-acceleration is mainly driven by the Wolf-Rayet star, WR20a. 

Besides the qualitative and quantitative analysis of the shell, we spectrally resolve and present the spatially distinct gas structures in RCW~49: the ridge and the northern and southern clouds.

Comparing our findings with the existing literature, we conclude that a secondary generation of star formation has been triggered in the shell but the new generation of stars being formed are relatively lower in mass than those existing in Wd2. \\

\acknowledgments
We thank the anonymous referee for bringing important issues to our attention and for helping to clarify the paper.

This work is based on observations made with the NASA/DLR Stratospheric Observatory for Infrared Astronomy (SOFIA). SOFIA is jointly operated by the Universities Space Research Association, Inc. (USRA), under NASA contract NNA17BF53C, and the Deutsches SOFIA Institut (DSI) under DLR contract 50 OK 0901 to the University of Stuttgart. Financial support for the SOFIA Legacy Program, FEEDBACK, at the University of Maryland was provided by NASA through award SOF070077 issued by USRA.

The FEEDBACK project is supported by the Federal Ministry of Economics and Energy (BMWI) via DLR, Projekt Number 50 OR 1916 (FEEDBACK) and Projekt Number 50 OR 1714 (MOBS - MOdellierung von
Beobachtungsdaten SOFIA).

This work was also supported by the Agence National de Recherche (ANR/France) and the Deutsche Forschungsgemeinschaft (DFG/Germany) through the project ``GENESIS” (ANR-16-CE92-0035-01/DFG1591/2-1).

\appendix

\section{Improved Baseline reduction algorithm using a Principal Component Analysis}\label{sec:pca_appen}

A general description of PCA technique and its application in astrophysics can be found in \citet{1997ApJ...475..173H} and \citet{1997ApJ...482..245U}. We employ this novel method to produce upGREAT spectra with a higher quality than is possible with a standard polynomial baseline removal.
To do so we identify systematic variations of the baseline between multiple spectra of the same receiver element. These variations are caused by instabilities in the telescope system, i.e. backends, receiver, telescope (optics), and atmosphere over the course of the observations. In the calibration step we produce additional data from the OFF-source (emission-free background) measurements of the on-the-fly data by subtracting subsequent OFF positions from each other. We calibrate these ``OFF-OFF" spectra in the same way the ``ON-OFF" spectra are calibrated. This additional set of spectra contains all the dynamics of the telescope system and the atmosphere without the astronomical information. By the use of a Principal Component Analysis (PCA) we identify systematic ``components" or ``eigenspectra'' that explain most of the variance away from the mean between these spectra. This is done separately for each of the 14 receiver elements of the upGREAT array and for each flight in which the source was observed. Using a linear combination of the strongest ``components" we try to describe the ``ON-OFF" spectra as best as possible by finding the best fit coefficients for each component. Subsequently we scale each  component by the coefficients that we found and subtract them from the ``ON-OFF" spectra. This removes the systematic variations found in the ``OFF-OFF" spectra, but does not alter the astronomical information in the ``ON-OFF'' spectra. With this technique we can correct very complex baseline features that are difficult or impossible to correct with the standard polynomial baseline removal. We are currently preparing a paper that describes this method in more detail (Buchbender et al. in prep.).

\clearpage

\section{Expansion of the shell}\label{sec:shell_exp_appen}
\label{rb_cii}
\begin{figure}[h!]
\centering
\includegraphics[width=70mm]{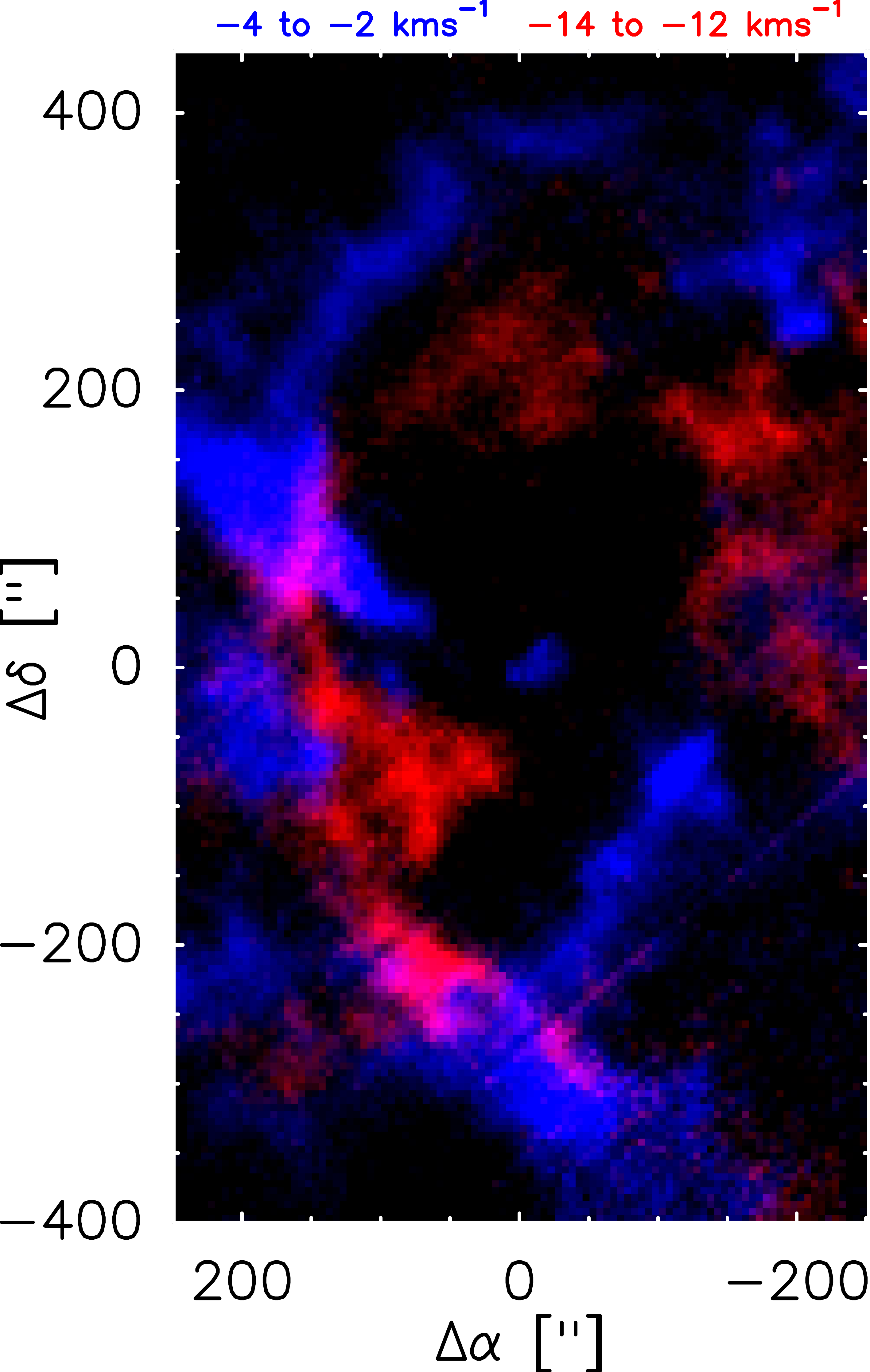}
\caption{Different channels of \cii\ emission in the velocity range of -14 to -12~km~s$^{-1}$ (red) and -4 to -2~km~s$^{-1}$ (blue), depicting the expansion of the shell as seen in Fig.~\ref{cii-chan}.}

\label{rb_-12_2} 
\end{figure}

\clearpage

\section{PV diagrams}\label{sec:pv_appen}

\begin{figure}[h]
    \centering
    \includegraphics[width=\textwidth]{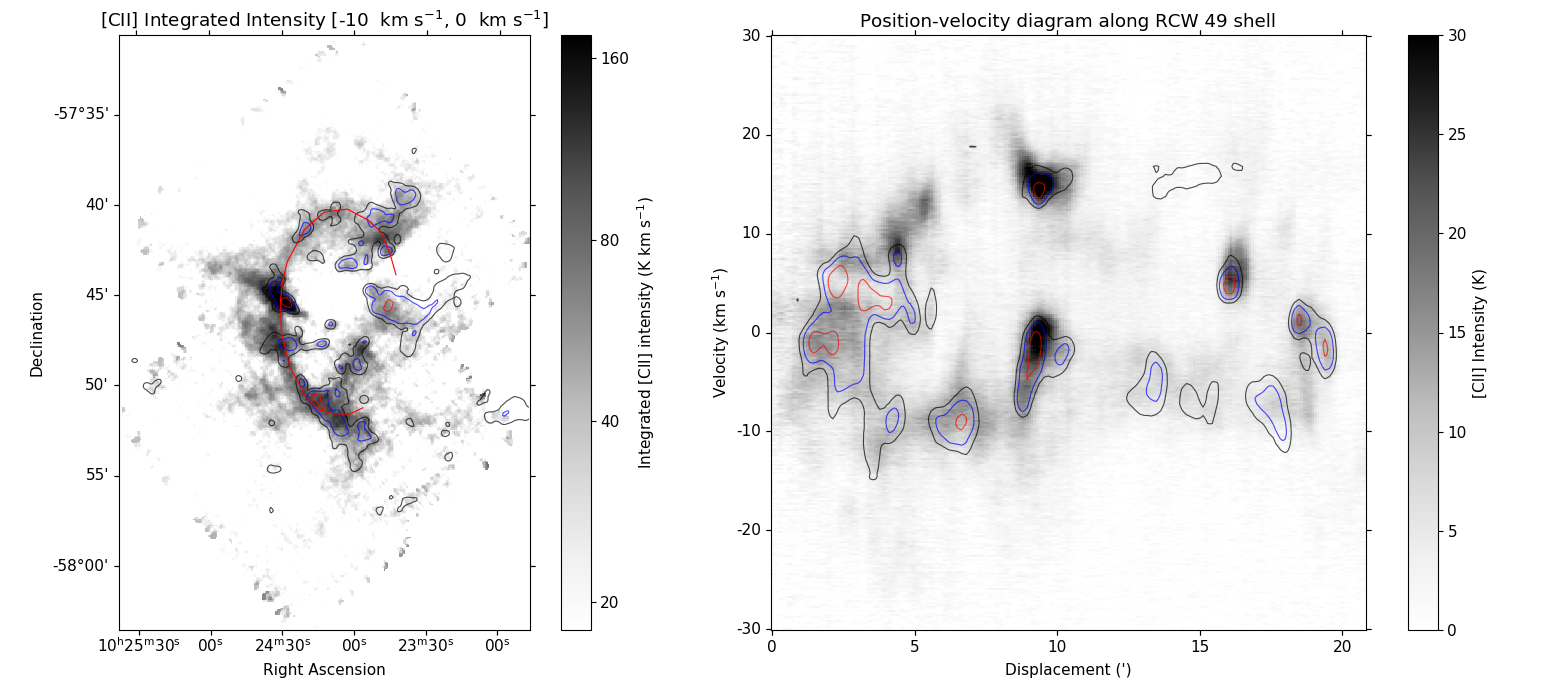}
    \caption{\textit{(Left)} Integrated \cii\ intensity from $-10$ to 0~km~s$^{-1}$ in greyscale. Integrated $^{12}$CO(3$-$2) within the same velocity interval in contours, which mark integrated intensities of 20, 50, and 125~K~km~s$^{-1}$ in black, blue, and red contours respectively. The overlaid red curve tracing the limb-brightened shell marks the path for the pv diagram in the right hand plot. \textit{(Right)} The pv diagram along the red path in the left hand plot, with 0$\arcmin$ displacement at the southern end of the red curve. As in the right hand plot, \cii\ in gray-scale and $^{12}$CO(3$-$2) in contours, with black, blue, and red contours marking intensities of 4, 8, and 16~K respectively. In both the integrated intensity and pv diagram, $^{13}$CO(3$-$2) generally follows the $^{12}$CO.}
    \label{fig:shell_pv}
\end{figure}

\clearpage

\begin{figure}[htp]
\centering
\includegraphics[width=140mm]{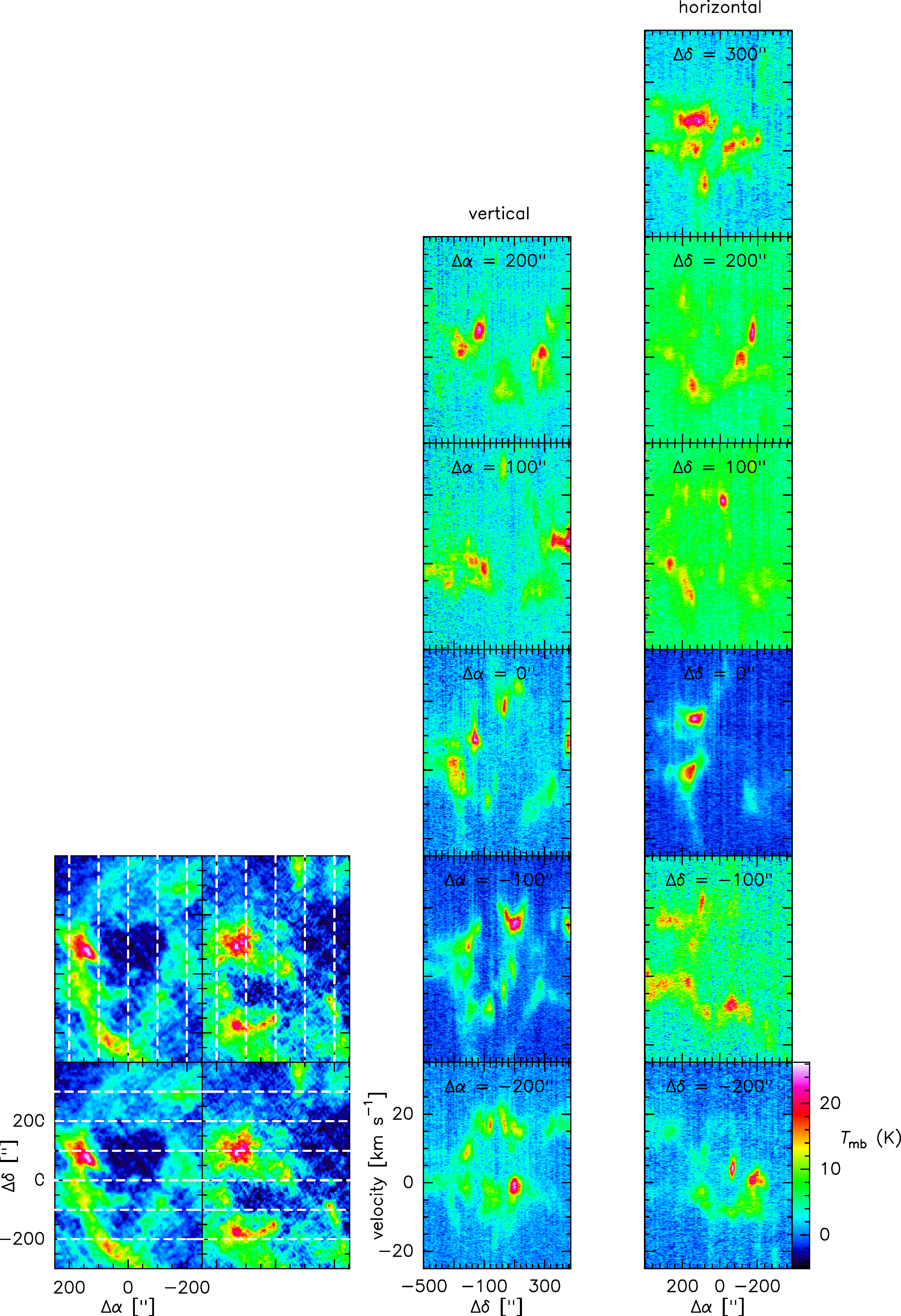}
\caption{The first two columns show the \cii\ velocity integrated intensity maps in the range from $-$25 to 0~km~s$^{-1}$ and from 0 to 30~km~s$^{-1}$, respectively. They are marked with vertical and horizontal cuts (white dashed lines), along which the pv diagrams are shown in columns named vertical and horizontal, respectively. The color bar is common for all panels.} 

\label{pv-ver} 
\end{figure}

\section{Optical depth of \cii\,}\label{sec:cii_opacity_appen}

We determined the opacity of \cii\ emission using its isotope \13cii\,. The \13cii\ line splits into three hyperfine-structure (hfs) components due to the coupling of angular momentum and spin of the hydrogen nucleus. Due to the limited signal-to-noise (S/N) ratio we have toward any single line-of-sight, we averaged the spectra toward a bright region (100$\arcsec$ $\times$ 100$\arcsec$ around $\Delta\alpha$ = 150$\arcsec$, $\Delta\delta$ = 80$\arcsec$) in \cii, emission, to detect \13cii\ lines. We used the $F$ = 1 $\to$ 0 hfs component at 1900.95~GHz, which is the second strongest hfs component with a relative intensity ($r_{\rm i}$) of 0.25 (see \citealt[Table.~1]{2020arXiv200212692G}), to calculate the total \13cii\ intensity. The reason we did not use the brightest hfs component is because it lies within the velocity wing of the \cii\ emission, while the second strongest component lies well beyond the velocity range of \cii\ emission and can be analysed. The $F$ = 1 $\to$ 0 hfs component of \13cii\ is multiplied by the $^{12}$C/$^{13}$C ratio, $\alpha$ = 52 \citep{2005ApJ...634.1126M}, for a Galactocentric distance of $\sim$ 8~kpc for RCW~49 (using equation~2 of \citealt{1993A&A...275...67B}).  \\

\begin{figure}[htp]
\centering
\includegraphics[width=85mm]{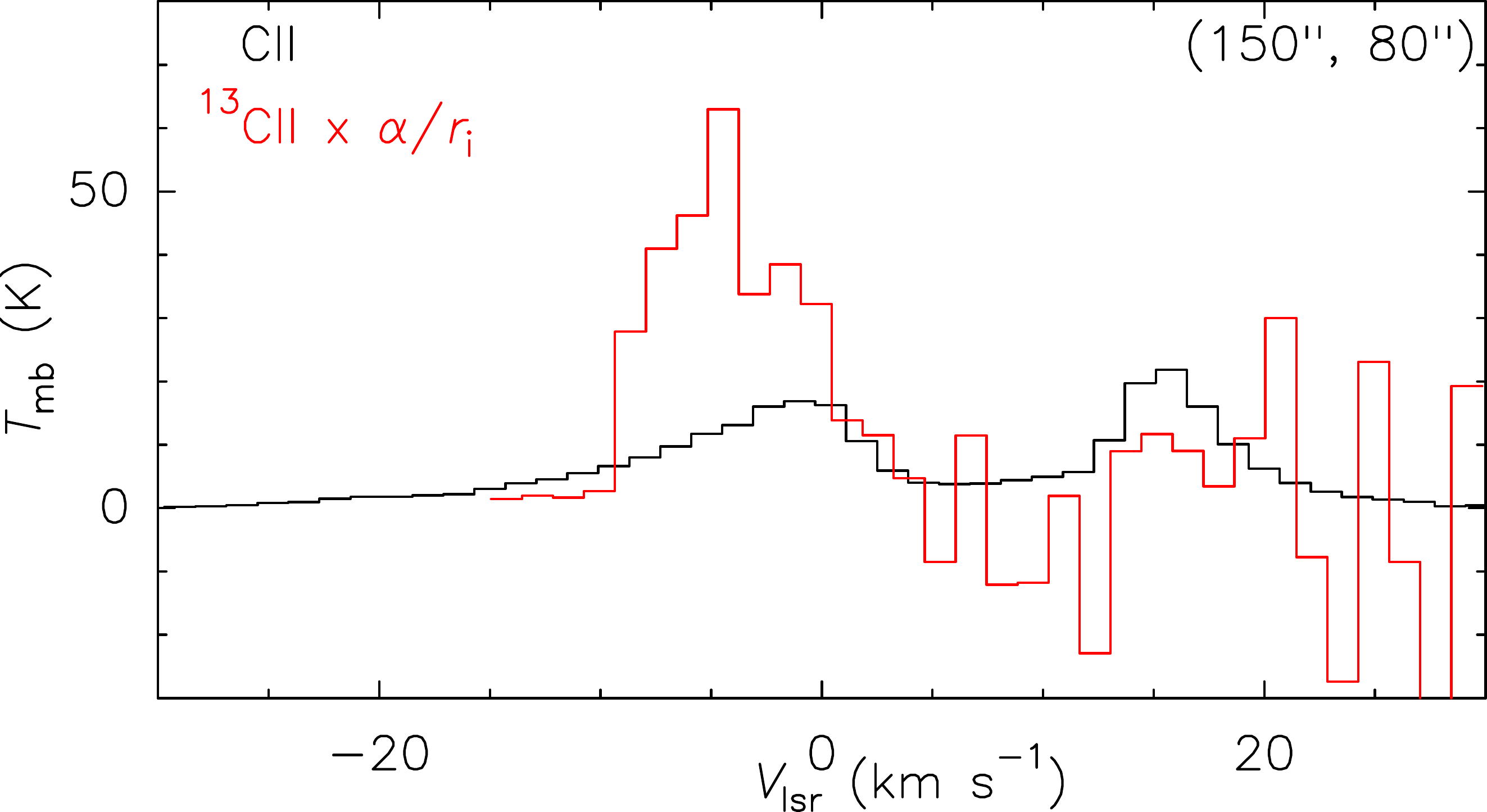}

\caption{Average spectra of \cii\ and \13cii\ toward a bright 100$\arcsec$ $\times$ 100$\arcsec$ area around ($\Delta\alpha$ = 150$\arcsec$, $\Delta\delta$ = 80$\arcsec$). The $^{12}$C/$^{13}$C ratio is denoted by $\alpha$ and $r_{\rm i}$ is the relative intensity of $F$ = 1 $\to$ 0 hfs component of \13cii\,.}

\label{cii-13cii-spec} 
\end{figure}

As can be seen in Fig.~\ref{cii-13cii-spec}, we find that the \13cii\ spectral emission follows a similar profile as that of \cii\, within its higher noise, but the intensity is higher than expected for optically thin \cii\,, indicative of an optical depth $>$ 1. Using the technique mentioned in \citet{2020arXiv200212692G} and using their equation~4, we estimated an optical depth in \cii\,, of $\tau$ = 3. This number should only be taken as a reference because we find that for an rms of 0.8~K, we get a \13cii\ peak detection of $\sim$ 1.2~K i.e. a S/N of 1.5. Due to reduced S/N, we were unable to estimate \cii\ optical depths toward other regions. So, we take $\tau$ = 3 as an upper limit for the entire \cii\ emission toward RCW~49. \\

\section{CO Excitation temperature and column density}\label{sec:co_tex_colden_appen}

Assuming $^{12}$CO to be optically thick and that both $^{12}$CO and $^{13}$CO have same excitation conditions under LTE, we can adapt the formalism as described in \citet{2018A&A...615A.158T} to determine $T_{\rm ex}$. Using $J$ = 3 $\to$ 2 transition of $^{12}$CO, we can calculate $T_{\rm ex}$ by

\begin{equation}
    T_{\rm ex} = 16.6~\bigg[\rm ln\bigg(1 +\frac{16.6}{T_{\rm MB}(^{12}\mathrm{CO})}\bigg)\bigg]^{-1} K\,, 
\end{equation}

\noindent where $T_{\rm mb}$ is the main beam brightness temperature and the constant ($h\nu/k$ = 16.6~K) is calculated for $\nu$ = 345.769~GHz, which is the frequency for $J$ = 3 $\to$ 2 transition of $^{12}$CO. For instance, for an average $T_{\rm MB}$($^{12}$CO) within the masked region shown in Fig.~\ref{fig:dust_mask} is $\sim$ 8~K, we get an average $T_{\rm ex}$ $\sim$ 15~K. Using the pixel-by-pixel calculation of $T_{\rm ex}$, we can determine the $N$($^{13}$CO) using

\begin{equation}
    N(^{13}\mathrm{CO}) = 5.29 \times 10^{12} (T_{\rm ex} + 0.88)~ {\rm exp}\Bigg(\frac{31.7}{T_{\rm ex}}\Bigg)  \Bigg(\frac{\tau_{\rm 13}}{1-{\rm exp}(-\tau_{\rm 13})}\Bigg)  \int T_{\rm MB}(^{13}\mathrm{CO}){\rm d}v~{\rm cm^{-2}}\,.
\end{equation}

Here the constants are $3kQ_{\rm rot}/8\pi^3\nu\mu^2J_{up}$ = 5.29 $\times$ 10$^{12}$($T_{\rm ex}$ + 0.88) and the upper level energy, $E_{\rm up}$ = 31.7~K, determined for the partition function, $Q_{\rm rot}$ = 0.38$T_{\rm ex}$ + 1/3, the dipole moment, $\mu$ = 1.1 $\times$ 10$^{-19}$ esu and the upper level, $J_{\rm up}$ = 3. The optical depth of $^{13}$CO, $\tau_{\rm 13}$, can be calculated for an optically thick $^{12}$CO by:

\begin{equation}
    \tau_{\rm 13} = -{\rm ln}\Bigg[1 - \frac{T_{\rm MB}(^{13}CO)}{15.87} \Big(\frac{1}{{\rm exp}(15.87/T_{\rm ex})-1}-0.003\Big)^{-1}\Bigg].
\end{equation}

Here the constants are $h\nu/k$ = 15.873~K and 1/(exp($h\nu/kT_{\rm bg}$) - 1) = 0.003, calculated for $\nu$ = 330.588~GHz, which is the frequency for $J$ = 3 $\to$ 2 transition of $^{13}$CO and for a background temperature, $T_{\rm bg}$ = 2.75~K. For the column density estimation within the half-elliptical mask shown in Fig.~\ref{fig:dust_mask}, we used the average main beam brightness temperatures of $^{12}$CO and $^{13}$CO and determined an average $\tau_{\rm 13}$ = 0.325. 


\section{Westerlund 2 Catalog and Stellar Properties} \label{sec:wd2_appendix}

In order to estimate the potential influence of the known stellar population on its environment, we synthesized a catalog of early-type stars from the work of \citealt{TFT2007} (hereafter \citetalias{TFT2007}), \citealt{VA2013} (hereafter \citetalias{VA2013}), and \citealt{VPHAS_Wd2_2015} (hereafter \citetalias{VPHAS_Wd2_2015}).
From these three catalogs, we should collect most of the known O and early B stars associated with Westerlund 2.
We intended to collect as many early-type candidates as possible so that we could evaluate the feedback contribution of the few most massive stars compared to the contribution of the full catalog of suspected OB stars, so we accepted stars which fulfilled any of several relaxed constraints.
We admitted into our list of early-type candidates any stars 1) with known OB spectral types listed by \citetalias{VA2013} in their Table 6 or \citetalias{VPHAS_Wd2_2015}; or 2) marked by \citetalias{VPHAS_Wd2_2015} as `WD2' cluster candidates based on similar extinction values; or 3) marked by \citetalias{TFT2007} as `ET' candidates based on their NIR colors; or 4) present in Table 3 of \citetalias{VA2013} in sub-tables ``Stars with only absorption lines'' or ``Stars believed to be late O/early B.''
In order to cast a wide net for OB candidates, we only required stars to fulfill one of these criteria, though many fulfilled several.
\citetalias{VPHAS_Wd2_2015} identified objects in their catalog that were also cataloged by \citetalias{TFT2007} or \citetalias{VA2013}, and we did no further cross-matching of our own between the \citetalias{VPHAS_Wd2_2015} catalog and either of the other two.
For candidates from either \citetalias{TFT2007} or \citetalias{VA2013} that were not in the \citetalias{VPHAS_Wd2_2015} catalog, we cross-matched between those two catalogs.

We find a total of 83 massive stars associated with Wd2, though these include a few stars that \citetalias{VPHAS_Wd2_2015} identifies as potential runaways or ejectees due to their similar reddening to the rest of the cluster.
Of the 83 total massive stars, 66 are within 12$\arcmin$ of the cluster center, close enough to influence the thermal and kinematic properties of the \hii\ region; 60 are within 6$\arcmin$, and 50 are within 3$\arcmin$.
We did not explicitly check our synthesized catalog against those of \citet{2011A&A...535A..40R}
or \citet{Hur2015},
among others, so it's possible that we could be missing a small number ($\sim < 5$) of O or early B stars within $\sim$12'.
We are aware that \citet{Zeidler2018}, by their analysis of \textit{VLT}/MUSE spectra, add 2 new O stars (O7.5 and O8.5) and 5 new B stars to the list of known spectral types near the cluster center, which we have not included in our catalog.

We adopted OB spectral types first from \citetalias{VPHAS_Wd2_2015}, since it is the most recent work, and then from \citetalias{VA2013}.
For all remaining stars, we assigned the uncertain type O8--B1.5. \\

We used the theoretical calibrations of \cite{Martins2005} to estimate effective temperature $T_{\text{eff}}$, surface gravity log$~g$, and luminosity $L$ from spectral type.
For WR20a, we adopt the fitted parameters $T_{\text{eff}}$, $R_{*}$, $v_{\infty}$, and $\dot M$ from \cite{Rauw2005}, who has suggested that the two binary components are nearly identical based on their spectra.
We note a potential caveat here: \citet{Rauw2005} assumes a heliocentric distance close to 8~kpc.
We use only the four spectroscopically fitted parameters listed above, and it is not clear to us how the heliocentric distance affects those particular parameters in their analysis.
In any case, this is the only available measurement of these parameters which are necessary for our feedback capacity analysis of the WR stars.

While the spectrum of WR20b has not been modeled in such detail, the star has been assigned the same type (WN6ha; \citealt{vanderHucht2001}) as each component of WR20a (WN6ha+WN6ha; \citealt{Rauw2005}), so we adopt the same parameters for WR20b as for one component of WR20a.
Given the appropriate parameters, we picked out models for each star from the PoWR stellar atmosphere grids \citep{PoWRCode3}.

Each of the PoWR models provides a synthetic spectrum, from which we integrate the total ionizing flux between 6--13.6 eV in order to calculate $G_{0}$ from the ensemble of stars.
For each location in a coordinate grid covering the entire \hii\ region, we add up the ionizing flux from every star using projected distances assuming all stars lie in a plane at 4.16 kpc.
We can use this grid to find $G_{0}$ at any location throughout the region; at the location of the bright Eastern shell, $G_{0} \sim$ 2--3 $\times~10^{3}$ in Habing units.

For the total mass loss rate $\dot M$ of the cluster, we take the individual mass loss rates of the OB stars from \cite{Leitherer2010} using the $T_{\text{eff}}$ and log$~g$ calculated above.
For the WR stars, we use the mass loss rate from \cite{Rauw2005}.
We sum over all stars in the cluster to find the total mass loss rate.

Finally, we calculate the total mechanical luminosity $L_{\text{mech}}$ of the cluster by summing over $L_{\text{mech}}$ of each star. We calculate $L_{\text{mech}} = \frac{1}{2} \dot M v_{\infty}^2$ using the terminal wind velocities $v_{\infty}$ and mass loss rates $\dot M$ of each star from \citet{Leitherer2010} for OB and \citet{Rauw2005} for WR. \\

We take a statistical approach to determining the values and uncertainties of these cluster properties.
For each star, we create a set of possible spectral types including 1) an inherent half sub-type calibration uncertainty and 2) the stated range of possible spectral types, when present.
For the WR stars, we take stated uncertainties associated with the parameters assigned by \cite{Rauw2005} and sample from them random realizations of parameter combinations for the WR stars.
Uncertainty was not specified for mass loss rate, so we assume a 10\% inherent uncertainty in $\dot M$, though we primarily drive the mass loss rate uncertainty by scaling it with $\sqrt{f}$, where $f$ is the volume filling factor which we vary between 0.1--0.25 (\citealt{Rauw2005} assumes 0.1, while \citealt{PoWRWN2} assumes 0.25).

With these sets of possible spectral types or parameter combinations for each cluster member, we draw random realizations of the entire cluster.
For each realization, we assign FUV flux, mass loss rate, and mechanical luminosity as described above, sum them across the entire cluster realization, and then take the median of the summed values of all cluster realizations.
We use the 16th and 84th percentiles of these distributions for the lower and upper error bars, respectively.

\section{Comparison to \texttt{Starburst99}} \label{sec:sb99_appendix}

We augment our analysis of the energetics in Section~\ref{sec:energetics} with predictions made using \texttt{Starburst99}, a spectral synthesis software which accepts a cluster IMF description and simulates population synthesis and cluster evolution over time by following stellar evolution models \citep{Leitherer2014Starburst99}.
The software outputs stellar spectra, wind properties (which are of particular interest to us), and other synthesized cluster properties at desired evolutionary time steps.
We can compare our configuration with that of \citet{Rauw2007}, who used \texttt{Starburst99} to estimate the wind power of Westerlund~2 using a Salpeter IMF and a total mass of 4500~$M_{\odot}$ in stars of masses between 1 and 120~$M_{\odot}$.

All inputs to the software are left as default unless otherwise specified.
We base the cluster mass function properties on the IMF fitted by \citet{Ascenso2007}, with a slope of $\Gamma = -1.20 \pm 0.16$ and total mass 2809~$M_{\odot}$ in stars of masses between 0.8 and 11~$M_{\odot}$.
From these values, we calculate the total mass in stars between 1 and 100~$M_{\odot}$.
We adopt the lower limit of 1~$M_{\odot}$ from \citet{Rauw2007} and the upper limit of 100~$M_{\odot}$ by averaging the 80~$M_{\odot}$ upper limit from the most massive observed star \citep{Zeidler2017} and the $\sim120~M_{\odot}$ upper limit suggested by \citet{Ascenso2007} and used by \citet{Rauw2007}.

The uncertainty on the IMF slope has a significant impact on the derived total cluster mass in our adopted mass bin, so we estimate the uncertainty on the modeled cluster wind properties due to the IMF slope uncertainty.
We sample a large number ($N\sim 1000$) of IMF slope values from the distribution suggested by \citet{Ascenso2007}, $\Gamma = -1.20 \pm 0.16$, assuming they describe a Gaussian distribution with $\mu = -1.20$ and $\sigma = 0.16$, and use each value to independently calculate the total mass in stars between 1 and 100~$M_{\odot}$.
From this mass distribution, we take the median (4000~$M_{\odot}$) and 16th and 84th percentile values (2900 and 5700~$M_{\odot}$) as the value and lower and upper error bars, respectively.
In order to avoid running a large number of \texttt{Starburst99} simulations, we simply run three simulations with these three total mass values, and all other inputs kept as described above.
For all predictions made using \texttt{Starburst99}, we thus use the ``median simluation'' as the predicted value and the upper and lower ``error bar simulations'' as the error bars, as in Figure~\ref{fig:sb99}.
We expect the total cluster mass to have a larger impact on the predicted wind power output than the IMF slope, so we do not vary the IMF slope in the simulations themselves.

\bibliography{references1}{}
\bibliographystyle{aasjournal}

\end{document}